%% file: llr-frames.tex
\RequirePackage{fix-cm}
\documentclass[smallextended]{svjour3} 

\usepackage{graphicx}
\usepackage{amsmath}
\usepackage{breqn}
\usepackage{url}

\journalname{Journal of Geodesy}

\begin{document}

\title{Role of lunar laser ranging in realization of terrestrial, lunar, and ephemeris reference frames}

\titlerunning{Role of LLR in realization of terrestrial, lunar, and ephemeris reference frames}

\author{Dmitry Pavlov}

\institute{D. Pavlov \at
  Institute of Applied Astronomy of the Russian Academy of Sciences (IAA RAS) \\
  Russia, 191187, St. Petersburg, Kutuzova Embankment, 10 \\
  Tel.: +7 (812) 275-1118 \\
  Fax.: +7 (812) 275-1119 \\
\email{dpavlov@iaaras.ru}
}

\date{Received: date / Accepted: date} 

\maketitle

\begin{abstract}
  Three possible applications of lunar laser ranging to space geodesy
  are studied. First, the
  determination of daily Earth orientation parameters (UT0 and variation of
  latitude), which is rarely used nowadays in presence of all-year VLBI, SLR,
  and GNSS data. The second application is the determination of two (out of
  three) lunar orientation parameters, i.e. daily corrections to the rotational
  ephemeris of the Moon. It may be of importance for the future lunar satellite-based
  navigational systems. The third application is the tie of ephemeris frame
  (BCRF) to the ICRF. It has been studied before, though in this work it is
  extensively compared to another realization of the same tie, obtained by
  spacecraft VLBI observations; also, two different EOP series and two different
  models of tidal variations of geopotential are applied, with different
  outcomes on the tie.

  The EPM lunar-planetary ephemeris, along with its underlying dynamical model and
  software, was used to obtain the presented results. All available observations
  were processed, since the earliest made at the end of 1969 at the McDonald
  observatory till the end of July 2019 (Matera, Grasse and also Wettzell observatory
  which began to provide data in 2018). The results and some open questions are
  discussed.
\end{abstract}

\section{Introduction}\label{sec:intro}

Lunar laser ranging (LLR), being the most precise technique of observation of the
relative motion of two Solar system bodies, through 50 years of existence
has produced many scientific results. Some of them are dedicated to fundamental
properties of spacetime
(see e.g.~\cite{Hofmann2010,Williams2012,Williams2014,Viswanathan2018,Hofmann2018Relativistic})\footnote{A collection
  of materials on LLR and relativity is available at \url{http://www.issibern.ch/teams/lunarlaser}},
while some others have provided new information about
tides on the Earth~\cite{Williams2016}, and tides and the internal structure of the
Moon~\cite{Williams2015,Matsumoto2015,pavlov2016}. In addition to the theoretical results,
LLR has allowed to build a high-precision geocentric ephemeris of the Moon,
featuring both orbital and rotational motion~\cite{DE430TR,EPM2017,Viswanathan2018}. The building of such ephemeris is
usually accompanied by determination of the positions of lunar laser ranging
stations (and velocities for some of them), and the positions of the five lunar
retroreflectors. The lunar ephemeris and the positions of the lunar retroreflectors
thus form a most precise lunar
reference frame that can be used for navigation and orbit determination in
future lunar missions, as well as for improvement of currently achieved
theoretical results. Russian planned lunar lander Luna-25 will have a
retroreflector panel~\cite{NPKSPPretroreflectors} on it;
there are also proposals for placing next-generation
single corner cubes on the surface of the Moon~\cite{Currie2011,Preston2012,TuryshevRetroreflector,Araki2016}.

There are just a few LLR stations in the world. They operate
independently from each other and are constrained by light and weather
conditions and Moon's visibility. Hence, the events of LLR from
different stations on the same day are rare. Still, with one station operating
on one day, it is possible to determine two daily (nightly) corrections to
Earth rotation angles: UT0 and VOL (variation of latitude), which are
mathematically the same as daily corrections to the station's longitude and
latitude, respectively. These corrections brought a significant contribution to
the determination of the terrestrial pole and Universal time in 1970--1980s, see
e.g.~\cite{langley1981,dickey1985} or~\cite[pp. M-17--M-20]{iers1994}, and are
still calculated and used in JPL KEOF EOP series~\cite{KEOF2017}.

Known EOP series that combine terrestrial and celestial poles---namely, IERS
C04~\cite{bizouard17} and IERS Bulletin A~\cite[pp.~94--116]{iers2017}\footnote{Another
  EOP series that provides both terrestrial and celestial poles is JPL EOP2: \url{https://eop2-external.jpl.nasa.gov}.
This relatively new product was unknown to author at the time of writing.}
are constructed without using LLR data. However, the Moon has a much more stable orbit than the
artificial satellites used in SLR and GNSS observations. One reason
for that is the Moon experiencing smaller perturbations from
nonspherical gravitational potential of the Earth than do artificial satellites
at low orbits; another reason is that the solar pressure acceleration of the Moon
is very small and almost non-intermitting.
Also, unlike GNSS and VLBI, LLR observations do not suffer from
daily clock offsets. That, together with increasing precision and frequency of
LLR observations worldwide, suggests that the LLR might
be helpful in the future for EOP determination.

It is well known that the LLR observations are sensitive not
only to the Earth's equator plane, but also to Earth's orbit plane, due to
the Sun affecting the orbit of the Moon (see e.g. \cite{Williams2018}). That
makes LLR different from radio ranging observations of e.g. Mars orbiters which have an
accuracy of 55 cm at best \cite{Kuchynka2012}. Laser ranging to the distance
of Mars or Venus has not ever been done. The accuracy of 55 cm means sub-$\mu$as
sensitivity to the ecliptic plane at the distance of orbit of Mars, but
mere 18 mas sensivity to Earth orientation at best. On
the other hand, VLBI observations of quasars, widely used to determine the
orientation of the Earth's equator w.r.t. celestial frame, have quite low
sensitivity to the orbit of the Earth.

The sensitivity of LLR to both ecliptic and equator, among other
things, allowed to determine the value of obliquity of the ecliptic
$\epsilon_{\mathrm{J2000}} = 84381.406^{\prime\prime}$ in~\cite{Chapront2002}, which is still
included to this day into the IAU system of astronomical constants~\cite{Luzum2011}.

There have been attempts to determine the celestial pole, in the form of
corrections to nutation theory, via
LLR~\cite{Williams1995,Zerhouni2009,Biskupek2011,Hofmann2018Contributions}.
The results are not as accurate as the ones obtained by VLBI. Moreover,
there are two major problems with determining the celestial pole
by LLR:
\begin{enumerate}
\item The constant corrections to celestial pole orientation, while can be
  formally determined from the LLR, are not separable from the orientation of
  the Sun--Earth--Moon system, or the whole ephemeris frame, in the celestial
  frame.  If the ephemeris (especially Earth orbit around the Sun) is fixed in
  the analysis, it constrains the determined orbit of the Moon to have a certain
  orientation in the celestial frame.
\item Linear or periodic terms in celestial pole, while can be formally
  determined from LLR, are hard to separate from similar terms coming from the
  imperfection of the model of the lunar motion, specifically from Earth tides,
  lunar tides, or lunar core.
\end{enumerate}

A straightforward solution to both problems is to avoid processing LLR observations
alone, but always process them together with an EOP series that include
the celestial pole determined from VLBI observations. Thus the doubts about the
celestial pole should be eliminated and we will be able to treat the found
``corrections to celestial pole'' as corrections to ephemeris frame orientation
(for constant terms, item 1 above), or to the lunar theory (for linear or periodic
terms, item 2 above). In reality, however, things become more diffucult, which will be
shown in Section~\ref{sec:orientation}.

\section{Data}\label{sec:data}

\subsection{LLR observations}

All available LLR observations from late 1960 up to the end of July, 2019 were
used in the experiments.  Table \ref{tbl-observations} shows the number and
timespan of observations processed from each station.

\begin{table}
  \caption{Lunar laser ranging observations.}
  \begin{tabular}{ | l | l | r | }
    \hline
    \textbf{Station} & \textbf{Timespan} & \textbf{\# of normal points} \\ \hline
    McDonald, TX, USA  & 1969--1985 & 3604 \\ \hline
    Nauchny, Crimea, USSR & 1982--1984 & 25 \\ \hline
    MLRS1, TX, USA     & 1983--1988 &  631 \\ \hline 
    MLRS2, TX, USA     & 1988--2013 & 3653  \\ \hline 
    Haleakala, HI, USA & 1984--1990 &  770  \\ \hline 
    Grasse, France (Ruby laser) & 1984--1986 & 1188 \\ \hline 
    Grasse, France (YAG laser)  & 1987--2005 & 8324 \\ \hline 
    Grasse, France (MeO green laser) & 2009--2019 & 1930 \\ \hline
    Grasse, France (infrared laser)  & 2015--2019 & 4762 \\ \hline
    Matera, Italy              & 2003--2019 & 233 \\ \hline
    Apache Point, NM, USA      & 2006--2016 & 2648 \\ \hline
    Wettzell, Germany          & 2018--2019 & 42 \\ \hline
    \textbf{Total}             & \textbf{1969--2019} & \textbf{27810} \\ \hline
  \end{tabular}
  \label{tbl-observations}
\end{table}
 
Apache Point Observatory observations~\cite{murphy12,murphy13} were downloaded from the APOLLO
website\footnote{\url{http://physics.ucsd.edu/~tmurphy/apollo/norm_pts.html}}.
Observations for McDonald/MLRS1/MLRS2~\cite{shelus}, Haleakala~\cite{HaleakalaLLR},
and Grasse (Ruby and YAG~\cite{samain} lasers) were
downloaded from the Lunar Analysis Center of Paris Observatory
(POLAC\footnote{\url{http://polac.obspm.fr/llrdatae.html}}).
Green and infrared~\cite{GrasseInfrared} Grasse
observations 2009--2018 were downloaded from Geoazur
website\footnote{\url{http://www.geoazur.fr/astrogeo/?href=observations}};
they also appeared
on the POLAC website some time later. Grasse observations from 2019 were kindly
provided by Jean-Marie Torre. Matera Laser Ranging Observatory observations up to 2015 were downloaded
from POLAC, while newer ones, since 2017 were obtained from
CDDIS\footnote{\url{ftp://cddis.gsfc.nasa.gov/pub/slr/data/npt_crd}},
as well as the Wettzell
observations~\cite{Wettzell}. The Crimean observations were recently found on a
shelf~\cite{CrimeaLLR}. Normal points were made from them by James Williams and
Dale Boggs~\cite{WilliamsPrivate}. Both raw observations and normal points
are publicly available\footnote{\url{http://iaaras.ru/en/dept/ephemeris/observations}}.
They do not have a
notable impact on the results and were included for the sake of history.

Each normal point contains a time of firing (UTC), signal delay due to range in UTC
seconds, and uncertainty. Some of the given
uncertainties were re-weighted before processing.  In particular, the
uncertainties for selected Matera observations in 2010--2012, with given values
below 1 picosecond, were treated as a result of a human error in decimal
exponent and scaled 1000x. Uncertainties of Apache Point observations made in
2006--2012 were scaled 2x--6x, as recommended on the APOLLO website.  Given
uncertainties for Grasse observations since September 1999 have not been
normalized to $1/\sqrt{N-1}$, where $N$ is the number of returned photons forming
the normal point; the normalization was applied before processing.

Selected groups of older (pre-2000) observations were scaled up 1.2x--1.9x
to match the postfit weighted root-mean-square (wrms). See~\cite{pavlov2016}
for details.

\subsection{Planetary observations}

VLBI observations of a spacecraft orbiting a planet when it passes near a known
radio source are essential for determining a tie between ephemeris frame and
ICRF (see Sec.~\ref{sec:orientation}). They are also known as $\Delta$DOR
observations.  The result of a session of such observations is the angular
position of planet w.r.t. the ICRF, either as a one-dimensional projection (in
case of single-baseline observations), or two-dimensional astrometrical
position (in case of a multiple-baseline observations on Very Long Baseline Array, VLBA).
Almost all the data
were taken from the webpage of Solar system dynamics (SSD) group at NASA
JPL\footnote{\url{https://ssd.jpl.nasa.gov/?eph_data}}, including single-baseline data
obtained from Phobos-2~\cite{Hildebrand}, Magellan, Galileo, Venus
Express~\cite{DE423IOM}, Mars Reconnaissance Orbiter (MRO), and Mars Odyssey;
and also astrometric positions of Saturn obtained from Cassini (in
\cite{Jones2014} there are three more observations not present at the website).
Astrometric positions of Mars obtained from MRO and Odyssey observations on VLBA
were taken from~\cite{Park2015}.

Orbits of the planets are best determined (in the ephemeris frame) from ranging data or spacecraft
orbiting Mercury, Venus, Mars, Jupiter, and Saturn, and also from ranging and
differenced range observations of Mars landers. Also, older planetary ranging data were
used for Mercury and Venus. Optical observations of natural satellites of
Jupiter and Saturn from different observatories were used to help determine
their planets' orbits.

Most of the data was taken from the aforementioned SSD webpage, including
ranging data obtained from MESSENGER~\cite{Park2017} and Juno; ranging data
obtained from Mars Global Survayor, Mars Odyssey, and MRO~\cite{Konopliv2011};
ranging and differenced range data obtained from Viking~\cite{Yoder1997} and
Pathfinder~\cite{Folkner1997}, and older radar and optical observation data.
Mars Express and Venus Express ranges~\cite{MorleyBudnik} were downloaded
from the Geoazur website\footnote{\url{http://www.geoazur.fr/astrogeo/?href=observations/base}}.
Radar ranging data from Crimea is available on the
IAA RAS website\footnote{\url{http://iaaras.ru/en/dept/ephemeris/observations}}.

Processing of planetary ephemeris data, apart from VLBI, is not a topic of this
work; we refer the reader to other papers devoted to planetary ephemeris
\cite{DE430TR,Pitjeva2013,PitjevaPitjev2014}.

\subsection{Earth orientation parameters}

IERS C04 EOP series was used for processing of planetary
observations. For the processing of LLR observations, both C04
  and IERS Bulletin A weekly (``finals.all'') were used. While there are similarities between
  those series, they do not generally agree below 1 mas for a number of reasons:

\begin{itemize}
\item C04 is a ``long-term'' series and contains only past data with the lag of up
  to one month, while Bulletin A is ``rapid'' series, containing data up to present and a prediction.
  Also, Bulletin A is routinely retroactively updated upon the availability of new data.
  This is not the case with C04, except when it is done after a special decision\footnote{\url{ftp://hpiers.obspm.fr/iers/eop/eopc04/updateC04.txt}}.
\item C04 and Bulletin A are produced by different groups with different software,
  and, while underlying models are the same in general, there can be subtle differences
  in implementation.
\item Both series are calculated from a combination of VLBI, GNSS, and SLR data;
details of the combination process differ between the two series~\cite{bizouard17,iers2017}.
  \end{itemize}
  
The data were downloaded from the IERS
website\footnote{\url{https://www.iers.org/IERS/EN/DataProducts/EarthOrientationData/eop.html}}.

\section{Software}

ERA (Ephemeris Research in Astronomy), version 8 was used for processing the
planetary and lunar observations, refining the parameters and integrating the
dynamical equations \cite{PavlovSkripnichenko}. ERA is based on the Racket
programming platform \cite{plt-tr1,Racket} and has
SQLite\footnote{\url{http://sqlite.org}} as the database engine.

SOFA library\footnote{\url{http://www.iausofa.org}}~\cite{hohenkerk} was used for
conversion between terrestrial and celestial coordinates and conversion between
various time scales.

For numerical integration, an implementation of Adams--Bashforth--Moulton
method, modified to handle delay differential equations~\cite{DDE}, was used.
Time-delayed terms appears in the differential equations of lunar rotation
  due to the nature of the tidal dissipation.

\section{Models}\label{sec:models}

\subsection{Dynamical model}\label{sec:dynamical-model}

A single dynamical model of the Solar system (including Moon, planets,
asteroids, and Trans-Neptunian objects, TNOs), which serves as the basis for EPM
planetary-lunar ephemeris, was used in this work. The planetary part of the
model comprises relativistic accelerations of point-masses of the Sun, planets,
the Moon, asteroids and TNOs, as well as additional accelerations from solar
oblateness and Lense-Thirring effect. The orbital motion of the Earth is also
affected by ``point mass--figure'' accelerations that come from Sun, the Moon,
Venus, Mars, and Jupiter. The lunar part of the model comprises similar ``point
mass--figure'' accelerations, as well as torques, and a degree-2
``figure-figure'' torque between Earth and Moon. The tidal variations of the gravitational potential of
the Earth are taken into account, as far as the orbital motion of the Moon is
concerned~\cite{Williams2016}. Both orbital and rotational motion of the Moon are affected by the
rotational and tidal dissipation modeled as variations of the lunar
gravitational field and inertia tensor. For more detailed description,
we refer to \cite{pavlov2016} and \cite{Pitjeva2013,Pitjeva2018} for the lunar
and planetary parts, respectively.

One piece of lunar dynamical model particularly important for this work is the
model of tidal variations of Earth's grativational potential that come from the
Moon and the Sun raising periodical ocean and solid tides on the Earth. The
orbit of the Moon is perturbed by those variations. There are actually two
models that are used interchangeably; they were compared in~\cite{pavlov2016}:

\begin{itemize}
\item IERS2010 geopotential variations model~\cite{iers2010} with fixed
  coefficients and amplitudes. Only corrections up to order and degree 2 are
  taken into account.
\item DE430 model~\cite{Williams2016,DE430TR} of direct perturbing acceleration
  of the Moon. It comes with five time delays for three frequencies and three
  fixed Love numbers for those frequencies. Of the five time delays, three are
  fixed ``orbital'' delays applied when the positions of the perturbing objects (Sun
  and Moon) are calculated; two rotational delays (for diurnal and semi-diurnal
  frequencies) are determined parameters.
\end{itemize}

\subsection{Reductions}
Usual routines were applied during the processing of astronomical observations.
IAU2000/2006 precession-nutation model~\cite{WallaceCapitaine} together with IERS
EOP series~\cite{bizouard17} was used for transformation between terrestrial and
celestial reference frames. Relativistic delay of signal was calculated
according to theoretical approximation~\cite{kopeikin}; delays from the Sun,
Earth, the Moon, Jupiter, and Saturn were added up. Calculation of tropospheric
delay of laser signal was performed according to empirical models: zenith
delay~\cite{mendes04} and mapping function~\cite{mendes02}. As for radio
observations, the tropospheric delay was already substracted from the provided
data. Displacements of reference points due to solid Earth tides and pole tides
were calculated according to the IERS Conventions
2010~\cite{iers2010}.  Ocean loading was calculated from HARPOS files obtained
from the International Mass Loading
Service\footnote{\url{http://massloading.net}}~\cite{PetrovMassLoading}. Atmospheric
loading was not applied.

\subsection{Lunar orientation parameters}

In~\cite{pavlov2016}, the following lunar-to-celestial transformation matrix is
formulated:

\begin{equation}\label{eq:l2c}
  R_{\mathrm{L2C}} = R_z(\phi)R_x(\theta)R_z(\psi+\Lambda),
  \end{equation}

\noindent where $\phi$, $\theta$, and $\psi$ are Euler angles and are part of
the state of the dynamical system (see Sec.~\ref{sec:dynamical-model}), while
$\Lambda$ is a sum of three small kinematic periodic terms unaccounted for in the
dynamical model.

We treat $R_z(\phi)R_x(\theta)$ part in eq.~\ref{eq:l2c} as the lunar celestial
pole, and $R_z(\psi+\Lambda)$ as the lunar counterpart of the Earth rotation angle.  Also,
we assume that the Moon, like the Earth, experiences stochastic
variations in its rotation and in the position of instantaneous axis of rotation
in the lunar mantle (lunar pole):

\begin{equation}\label{eq:l2c-lop}
  R_{\mathrm{L2C}}^* = R_z(\phi)R_x(\theta)R_z(\psi + \Lambda)R_{z}(-r')R_y(q')R_x(p')
  \end{equation}

We call $p'$, $q'$, and $r'$ the lunar orientation parameters (LOP). $p'$ and $q'$
define the position of the lunar pole, while $r'$ is similar to Earth's
(UT1$-$UTC) correction.

While it is theoretically possible to further extend (\ref{eq:l2c-lop}) with the
corrections to the lunar celestial pole, there will not be a practical utility
in it before the development of astronomical instruments that are sensitive to
the position of the lunar celestial pole. There are propositions for such
instruments, e.g. lunar polar optical telescope~\cite{Petrova2012} or lunar
VLBI station~\cite{Kurdubov2018}. Even after they are developed, it may happen
that the present dynamical model already provides good enough lunar celestial
pole and that the corrections are not needed. Presently, with only LLR data
available, we restrict LOP to just $p'$, $q'$, and $r'$.

\subsection{Celestial pole}

In ephemeris--ICRF tie determination (Sec.~\ref{sec:orientation}), the model of
ITRS-to-GCRS is the IAU2000/2006 precession-nutation model which is extended with
two additional rotations:

\begin{align}\label{eq:t2c}
  R_{\mathrm{T2C}}(t) & = R_x\left(\Delta X\right) R_y\left(\Delta Y\right)
  Q(t)R(t)W(t) \nonumber\\
  \Delta X & = \Delta X_0 + \dot{X}(t - t_0) + C^{\Omega}_X \cos \Omega + S^{\Omega}_X \sin \Omega\\
  \Delta Y & = \Delta Y_0 + \dot{Y}(t - t_0) + C^{\Omega}_Y \cos \Omega + S^{\Omega}_Y \sin \Omega \nonumber
  \end{align}

where:
\begin{itemize}
\item $W(t)$ is the polar motion matrix, obtained from terrestrial pole from EOP series
  and adjusted for diurnal and semi-diurnal variations due to zonal and ocean tides;
\item $R(t)$ is the Earth rotation matrix, obtained from UT1 from EOP series
  and adjusted for diurnal and semi-diurnal variations due to zonal and ocean tides;
\item $Q(t)$ is the celestial pole matrix, adjusted by $\mathrm{d}X$ and $\mathrm{d}Y$
  corrections from EOP series;
\item $t_0$ is the epoch J2000; 
\item $\Omega$ is the moon ascending node, presessing with a period of approximately 18.6 years;
\item $\Delta X$ and $\Delta Y$ are determined constant rotations of current ephemeris frame to the ICRF;
\item $\dot{X}$ and $\dot{Y}$ are determined rotational trends of the current ephemeris frame w.r.t. the ICRF;
\item $C^{\Omega}_X$, $S^{\Omega}_X$, $C^{\Omega}_Y$ and $S^{\Omega}_Y$ are additionally determined artifacts that can appear as a 18.6 year periodic motion of celestial pole w.r.t the ICRF.
  \end{itemize}

$\dot{X}$, $\dot{Y}$, $C^{\Omega}_X$, $S^{\Omega}_X$, $C^{\Omega}_Y$ and $S^{\Omega}_Y$ are supposed to be zero because
both the ICRF frame, as realized by eq.~\ref{eq:t2c}, and the ephemeris frame
are assumed to be inertial. However, there are difficulties with those
assumptions, see Sec.~\ref{sec:orientation}.

\section{Main solution}\label{sec:main-solution}
A large set of parameters of lunar and planetary models was fitted to data (see
Sec.~\ref{sec:data}) using the nonlinear weighted least-squares method.  Lunar
and planetary parameters were fit (determined) one after another in several
iterations until both solutions converged with a joint lunar-planetary
ephemeris, which then served as a basis for the results presented in further
sections.

\subsection{Lunar part}
The following parameters were fitted in the lunar solution:

\begin{itemize}
\item Geocentric position and velocity of the Moon at epoch;
\item Euler angles of lunar physical libration ($\phi$, $\theta$, $\psi$) and their derivatives at epoch;
\item Angular velocity of the lunar liquid core at epoch;
\item Gravitational parameter of the Earth--Moon system;
\item Ratios of undistorted lunar moments of inertia: $\beta = (C - A) / B$ and $\gamma = (B - A) / C $
\item Stokes coefficients of undistorted lunar gravitational potential:
  $C_{32}$, $S_{32}$, and $S_{33}$;
\item Lunar Love number $h_2$;
\item Lunar core flattening coefficient;
\item Lunar tidal delay;
\item Rotational delays $\tau_{\mathrm{1R}}$ and $\tau_{\mathrm{2R}}$ of Earth
  diurnal and semi-diurnal tides (only if DE430 model of tidal variations of
  geopotential is used; otherwise, fixed IERS formula is applied);
\item Amplitudes of $l'$ (365~d), $2l-2D$ (206~d) and $2F-2l$ (1095~d)
  kinematic terms;
\item Selenocentric coordinates of five retroreflectors;
\item Terrestrial coordinates of all LLR stations;
\item Velocities of McDonald/MLRS1/MLRS2 and Grasse stations\footnote{for other stations, due
    to a relatively short timespan of their LLR observations, the velocities were taken from GNSS
    solutions. See~\cite{pavlov2016} for details.};
\item 24 biases for chosen stations at chosen periods of time.
\end{itemize}

A more detailed description of the lunar parameters can be found
in~\cite{pavlov2016}. 

The weighted root-mean-square postfit residuals of LLR observations are given in
Table~\ref{tbl:residuals}. Some observations with unreasonably large residuals
were considered erroneous and were rejected from the solution. It is visible
that the DE430 tidal model gives better results than IERS2010 (partially thanks
to two adjustable parameters); that has already been noticed in~\cite{pavlov2016}.
IERS2010 does not have the accuracy needed to model the orbit
of the Moon at the centimeter level on time intervals exceeding 20 years;
without adjustable parameters, the solution distorts the lunar dissipation model to soak up the acceleration.

The other consequence from \ref{tbl:residuals} is that the ``finals'' IERS EOP series give systematically
better fits than the C04 series; that is unexpected, given that the LLR data
was not used in production of either series.

\begin{table}[h]
  \caption{Post-fit statistics of lunar solution. WRMS is one-way and given in cm. Note:
  for pre-1980 McDonald data, KEOF EOP series was used instead of C04 or ``finals''.} 
  \begin{tabular}{ | l | l | r | r | r | r | r | r |}
    \hline
    \multicolumn{1}{|c|}{\textbf{Station}} &
    \multicolumn{1}{c|}{\textbf{Timespan}} &
    \multicolumn{1}{c|}{\textbf{Used}} &
    \multicolumn{1}{c|}{\textbf{Rej.}} &
    \multicolumn{4}{c|}{\textbf{WRMS}} \\
    \cline{5-8} 
    & & & & \multicolumn{2}{c|}{DE430 tides} & \multicolumn{2}{c|}{IERS2010 tides} \\
    \cline{5-8} 
    & & & & \multicolumn{1}{c|}{C04} & \multicolumn{1}{c|}{finals} &
            \multicolumn{1}{c|}{C04} & \multicolumn{1}{c|}{finals} \\ \hline
    McDonald           & 1969--1985 & 3552 & 52  & 20.2 & 20.1 & 21.1 & 20.9 \\
    MLRS1              & 1983--1988 & 588  & 43  & 11.5 & 11.0 & 12.4 & 11.5 \\ 
    MLRS2              & 1988--2013 & 3224 & 429 &  3.7 & 3.4  & 4.3  & 4.1 \\ 
    Nauchny            & 1982--1984 & 25   &  0  & 11.1 & 11.1 & 11.2 & 11.2 \\ 
    Haleakala          & 1984--1990 & 751  & 19  &  6.6 & 5.8  & 7.0  & 6.0 \\ 
    Grasse (Ruby)      & 1984--1986 & 1109 & 79  & 17.7 & 16.9 & 18.5 & 17.7 \\
    Grasse (YAG)       & 1987--2005 & 8273 & 51  &  1.7 & 1.5  & 2.0  & 1.9 \\
    Matera             & 2003--2019 &  219 & 14  &  3.2 & 3.1  & 3.2  & 3.2 \\ 
    Apache             & 2006--2016 & 2632 & 16  & 1.50 & 1.50 & 1.77 & 1.78 \\
    Grasse (MeO green) & 2009--2019 & 1930 &  0  & 1.61 & 1.64 & 2.01 & 1.97 \\ 
    Grasse (infrared)  & 2015--2019 & 4761 &  1  & 1.30 & 1.25 & 1.45 & 1.40 \\
    Wettzell           & 2018--2019 &  42  &  0  & 1.06 & 1.08 & 1.27 & 1.37 \\ \hline
  \end{tabular}
  \label{tbl:residuals}
\end{table}

The selenocentric coordinate system, in which the coordinates of five
retroreflector panels are determined, is based on the principal axes (PA) of the
Moon's figure.  Table~\ref{tbl:retroreflectors} lists the positions of the five
retroreflectors and their formal errors for the best solution of the four
(``finals'' EOP and DE430 tidal model). These five points realise the most
precise lunar coordinate system to date. The given formal errors, however,
  may be too optimistic. Table~\ref{tbl:retroreflectors-comparison} shows that,
  while the differences between positions determined
  in the four obtained solutions are 15 cm or less, the differences with
  DE430~\cite[Table 6]{williams13} or INPOP17a~\cite{INPOP17aNotes} positions reach 2 m.

\begin{table}[h]
  \caption{Determined positions of five reference points in PA selenocentric coordinate system.}
  \begin{tabular}{ | l | r | r | r | c | c | c |}
    \hline
    \multicolumn{1}{|c|}{\textbf{Panel}} &
    \multicolumn{1}{c|}{\textbf{X}, m} &
    \multicolumn{1}{c|}{\textbf{Y}, m} &
    \multicolumn{1}{c|}{\textbf{Z}, m} &
    \multicolumn{1}{c|}{$\pmb\sigma$\textbf{X}, cm} &
    \multicolumn{1}{c|}{$\pmb\sigma$\textbf{Y}, cm} &
    \multicolumn{1}{c|}{$\pmb\sigma$\textbf{Z}, cm} 
    \\ \hline
    Apollo 11 & 1591967.619 & 690697.773 & 21004.477 & 3.1 & 2.0 & 0.8  \\
    Apollo 14 & 1652689.300 & -520999.332 & -109729.671 & 3.1 & 2.1 & 0.8  \\
    Apollo 15 & 1554678.256 & 98093.848 & 765006.089 & 3.0 & 1.9 & 1.7  \\
    Luna 17 & 1114291.065 & -781299.633 & 1076059.333 & 3.0 & 1.7 & 2.0 \\
    Luna 21 & 1339364.109 & 801870.501 & 756359.405 & 3.0 & 1.9 & 1.7 \\ \hline
  \end{tabular}
  \label{tbl:retroreflectors}
\end{table}

\begin{table}[h]
  \caption{For each of the five reference points:
      maximum pairwise distance between its positions in the four obtained solutions;
      maximum distance between its positions in the obtained solutions and its position in the DE430 solution;
      maximum distance between its positions in the obtained solutions and its position in the INPOP17a solution.}
  \begin{tabular}{ | l | c | c | c |}
    \hline
    \multicolumn{1}{|c|}{\textbf{Panel}} &
    \multicolumn{1}{c|}{\textbf{Max $\Delta$ btw solutions}} &
    \multicolumn{1}{c|}{\textbf{Max $\Delta$ DE430}} &
    \multicolumn{1}{c|}{\textbf{Max $\Delta$ INPOP17a}} \\ \hline
    Apollo 11 & 0.15 m & 2.1 m & 2.3 m \\
    Apollo 14 & 0.11 m & 2.0 m & 2.2 m \\
    Apollo 15 & 0.13 m & 1.8 m & 2.2 m \\
    Luna 17 & 0.15 m & 1.7 m & 2.2 m \\
    Luna 21 & 0.15 m & 1.7 m & 1.9 m \\ \hline
  \end{tabular}
  \label{tbl:retroreflectors-comparison}
\end{table}

\subsection{Planetary part}
In the planetary solution the following parameters were determined:

\begin{itemize}
\item Three angles of orientation with respect to ICRF;
\item Planetary orbital elements at epoch, including Pluto. For Earth, only
  the eccentricity, semimajor axis and longitude of the periapsis were determined
  to avoid correlations with the three ICRF angles;
\item Solar oblateness factor;
\item Gravitational parameters of the Sun, some individual asteroids and TNOs;
\item Total gravitational parameters of asteroids of C, S, M taxonomic classes,
  asteroid belt and Kuiper belt (apart from asteroids/TNOs whose
  masses were fixed to known values or determined individually);
\item Rotational parameters of Mars;
\item Parameters of Mercury topography;
\item Locations of Martian landers Viking 1/2 and Pathfinder;
\item Solar corona electron density factors (one per each solar conjunction, assuming a symmetric $1/r^2$ distribution)
\item Shifts to compensate calibration errors or clock offsets on Earth or in
  spacecraft;
\item Phase effects for optical observations of outer planets.
\end{itemize}

For details about the planetary part of EPM ephemeris, the reader is referred
to~\cite{Pitjeva2013,PitjevaPitjev2014,EPM2017,Pitjeva2018}.

The whole Solar system was oriented to ICRF via single-baseline and
multiple-baseline $\Delta$DOR observations. While multiple-baseline
observations are given as the direct astrometrical position $(\alpha, \delta)$
of a planet, the single-baseline ones are given as a single observable
$\Delta\theta$. That observable is a linear combination of differences
$(\Delta\alpha, \Delta\delta)$ between the astrometrical of the planet
and its position in the DE405 ephemeris:

$$\Delta\theta = \Delta\alpha\cos\gamma + \Delta\delta \cos\alpha \sin\gamma,$$
where $\gamma$ is the angle of VLBI baseline on plane of sky, relative to celestial equator.

\begin{figure}[p]
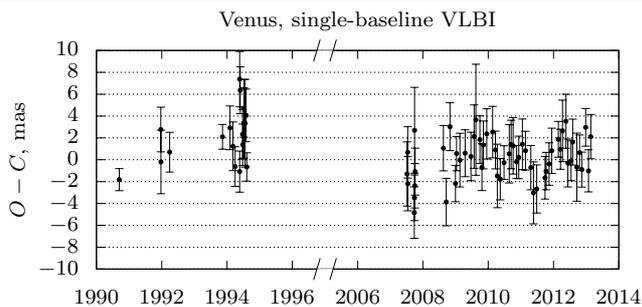

  \input vlbi-venus.tex
  \caption{Postfit residuals of single-baseline VLBI observable $\Delta\theta$ for Venus orbiters Magellan and Venus Express.}
  \label{fig:aldel-venus}
  \end{figure}

\begin{figure}[p]
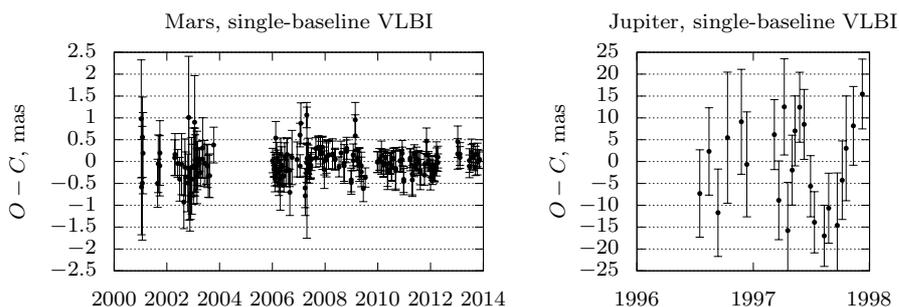

  \input vlbi-mars.tex
  \input vlbi-jupiter.tex
  \caption{Postfit residuals of single-baseline VLBI observable $\Delta\theta$ for Mars orbiters MGS, Odyssey, and MRO, and Jupiter orbiter Galileo}
  \label{fig:aldel-mars-jupiter}
  \end{figure}

\begin{figure}[p]
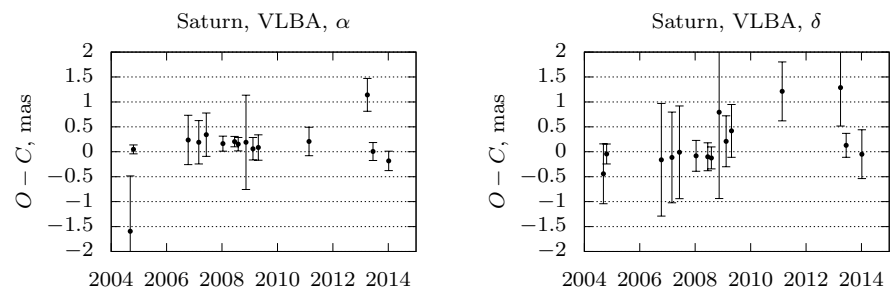

  \input vlba-saturn-alpha.tex
  \input vlba-saturn-delta.tex
  \caption{Postfit residuals of astrometric positions of Saturn obtained on VLBA observations of Cassini spacecraft.}
  \label{fig:vlba-saturn}
  \end{figure}

\begin{figure}[p]
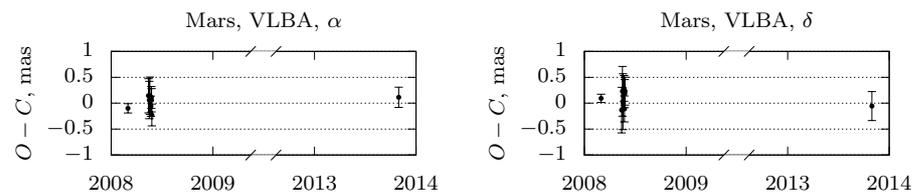

  \input vlba-mars-alpha.tex
  \input vlba-mars-delta.tex
  \caption{Postfit residuals of astrometric positions of Mars obtained on VLBA observations of MRO and Odyssey spacecraft.}
  \label{fig:vlba-mars}
\end{figure}

After orientation to ICRF, the following residuals were obtained:
Figs.~\ref{fig:aldel-venus} and~\ref{fig:aldel-mars-jupiter} for Venus, Mars,
and Jupiter single-baseline VLBI observables;
Figs.~\ref{fig:vlba-saturn} and~\ref{fig:vlba-mars} for Saturn and Mars
astrometric positions obtained by VLBA observations.

The formal errors (1$\sigma$) of the three angles of rotation (around X, Y, and
Z axes) of ephemeris frame to ICRF were the following: $\sigma(\Delta X) =
0.038$ mas, $\sigma(\Delta Y) = 0.041$ mas, $\sigma(\Delta Z) = 0.024$ mas.

The rotational trends ($\dot{X}$, $\dot{Y}$, $\dot{Z}$) were once temporarily
added into the set of the determined parameters, and were found to be below
their respective $1.5\sigma$, where $\sigma(\dot{X}) = 14$ $\mu$as/year,
$\sigma(\dot{Y}) = 15$ $\mu$as/year, and $\sigma(\dot{Z}) = 9$ $\mu$as/year. That
proves that the dynamical planetary model of the ephemeris properly accounts for
all natural phenomena that can have a rotational effect on the Solar system
within the given error margin. Estimates of the galactic aberration are below
that margin, at 5--10 $\mu$as/year~\cite{Kurdubov2010,Malkin2014}.

\section{Determination of daily corrections to rotational parameters of Earth and Moon}

An LLR session running for several hours allows to determine two
daily (nightly) parameters of Earth rotation: UT0 and VOL, which are in linear
relation with UT1 and terrestrial pole coordinates $x_p$, $y_p$~\cite{ChaprontFrancou}:

\begin{eqnarray*}
  \mathrm{UT0} &=& \mathrm{UT1} + \frac{(x_p \sin\lambda + y_p \cos\lambda)\tan\phi}{15\times 1.002737909} \\
  \mathrm{VOL} &=& x_p\cos\lambda - y_p\sin\lambda
  \end{eqnarray*}

\noindent where $\lambda$ and $\phi$ are the station's longitude and latitude,
respectively, and 1.002737909 is the relative rate of mean solar time to sidereal time.
Without the LLR sessions happening in two observatories on one night (which is
rare), there is no possibility to determine daily UT1, $x_p$, and $y_p$ from LLR
alone.

Lunar laser ranging is usually performed to more than one lunar target
(retroreflector).  That allows to determine two lunar orientation parameters:
$r'$ and $q'$. LLR is not sensitive to the third parameter, $p'$ --- the angle
of rotation around the X axis, directed towards the Earth.

\subsection{Setting of experiment}

Once the ephemeris, obtained with C04 as the EOP series, has been fixed, a
special LLR solution was obtained, containing five determined parameters for
each night with 10 or more LLR normal points: $\Delta$UT0, VOL, $r'$, $q'$, and
range error $\Delta r$. The range error is assumed to originate from
imperfections of the troposphere model and the lunar orbital ephemeris and
correlates with $\Delta$UT0 and VOL. The inclusion of the range error into the
set allows to obtain realistic error estimates for the EOP parameters.

\subsection{Results}

\begin{figure}
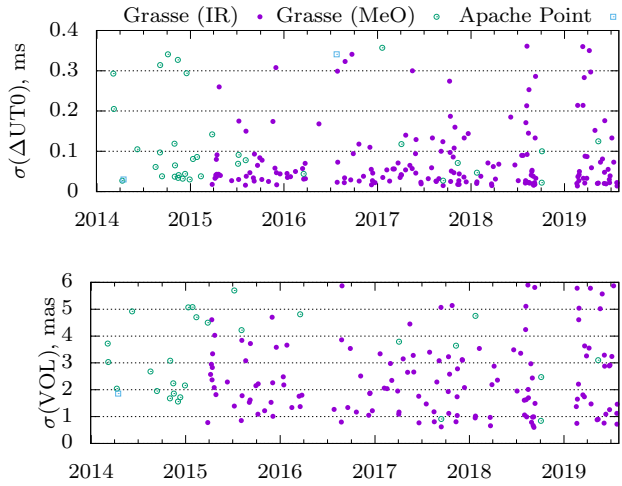

  \input ut0.tex
  
  \input vol.tex
  \caption{Formal errors of UT0 and VOL determined from LLR since 2014.}
  \label{fig:eop}
  \end{figure}

The results are shown for the timespan of January 2014 -- July 2019 for three
instruments at two observatories. There were 340 sessions altogether with 10 or
more normal points. There has not been such an LLR session at Wettzell
observatory in that timespan. On the plots, only the results with formal errors
(1$\sigma$) below 6 mas are shown. Fig.~\ref{fig:eop} shows the formal errors of
determined UT0 and VOL. 

\begin{figure}
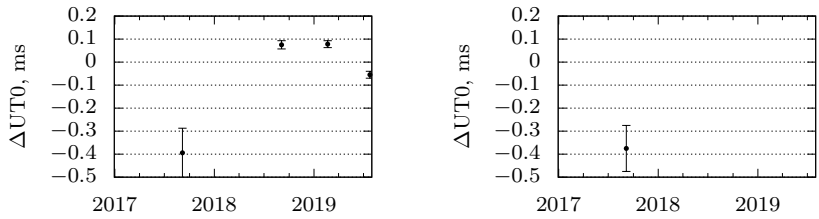

  \input ut0-corr-c04.tex
  \input ut0-corr-finals.tex
  \caption{$\Delta$UT0 values detected by Grasse infrared laser outside 3$\sigma$ w.r.t. C04 series (left) and ``finals'' (right).}
  \label{fig:ut0-corr}
  \end{figure}

There are 173 nights with $\sigma(\mathrm{\Delta UT0})$
below 1.5 mas (0.1 ms) and 50 nights with $\sigma(\mathrm{VOL})$ below
1.5 mas. Almost all determined corrections are within their $3\sigma$ range.
The outliers---four UT0 corrections w.r.t. C04 series and one
w.r.t. ``finals''---are shown at Fig.~\ref{fig:ut0-corr}. For C04, their
values and formal errors are:
$-394\pm 107 \mu$s (6 Sep 2017),
$75\pm 18 \mu$s (4 Sep 2018),
$78\pm 15 \mu$s (20 Feb 2019), $81\pm 16 \mu$s (4 Sep 2019), and
$-55\pm 15 \mu$s (25 Jul 2019). The only outlier for ``finals'' is
$-375\pm 100 \mu$s (6 Sep 2017).

The absense of the other three outliers in the UT0 corrections w.r.t. ``finals''
further supports the inference that the ``finals'' series represent the actual
rotation of the Earth slightly better than C04 at present and that the LLR can
sense it.

The single outlier common to both series can indicate either an Earth rotation
event not determined by daily VLBI observations, or an artifact in Grasse data
on that day. In any case, LLR will probably have a non-negligible effect on
modern IERS EOP solutions, and even more so when other LLR observatories, like
Wettzell, begin to provide frequent infrared data.

\begin{figure}
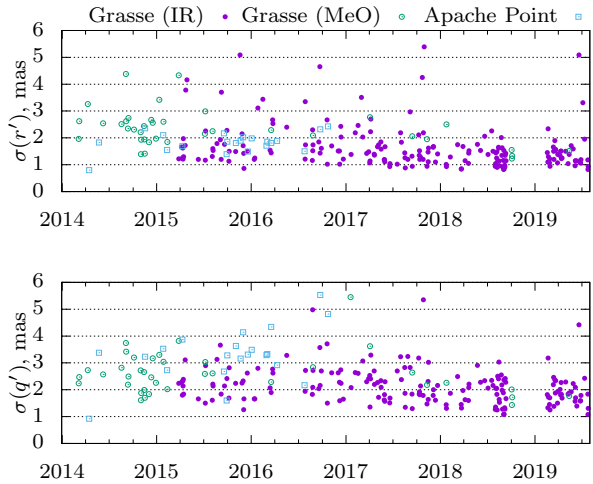

  \input lop-r.tex
  
  \input lop-q.tex
  \caption{Formal errors of $r'$ and $q'$ determined from LLR since 2014.}
  \label{fig:lop}
  \end{figure}

Fig.~\ref{fig:lop} shows the formal errors of $r'$ and $q'$ for nights where
those errors are below 6 mas. Data on Figs.~\ref{fig:eop} and~\ref{fig:lop} come
not from separate solutions but from single solution, where the range errors
were also determined. The range errors are not shown. All determined $p'$ and $q'$
are within their $3\sigma$ range.

\section{Determination of orientation of ephemeris frame in the ICRF from LLR data}\label{sec:orientation}

In planetary ephemerides, orbits of planets, including Earth, are determined in
a so-called barycentric celestial reference frame (BCRF~\cite{Kopeikin2011}), also known as
ephemeris frame. Like ICRF, it is defined to have its origin at the barycenter
of the Solar system. It is also defined to be an inertial frame. It is important
to tie dynamical BCRF to the kinematic ICRF as precisely as possible, for three
applications:
\begin{itemize}
\item Modeling the orbits of interplanetary spacecraft;
\item Studying the influence of the Moon, planets, and the Sun to the rotation of the Earth;
\item Supporting astronomical instruments on other planets and the Moon (in the future).
  \end{itemize}

Probably the earliest tie between the ephemeris and radio frame based on quasars
was made in 1986~\cite{Newhall1986}, long before the official adoption of ICRF
in 1998. The tie was based on $\Delta$DOR observations of Viking and Pioneer
spacecraft; the accuracy was about 20 mas. Later, the accuracy of the tie was improved
to about 5 mas~\cite{Finger92}, and then to 3 mas~\cite{Folkner94} by using a
different approach: processing LLR together with quasar VLBI observations.
However, in subsequent years, the abundance of a higher quality $\Delta$DOR
observations of spacecraft, including those of Magellan (since 1990),
quickly led back to the decision to use just $\Delta$DOR for the direct ephemeris-ICRF
tie~\cite{DE403}. In DE405 ephemeris~\cite{DE405} (1998), a 1 mas accuracy of the tie was
reported. Similar result was obtained in EPM2002 ephemeris~\cite{EPM2002}.

The modern ephemerides DE430~\cite{Folkner2012,DE430TR} and EPM2017~\cite{pitjeva2017,EPM2017}
have their tie to ICRF based on modern spacecraft VLBI measurements,
with accuracy of about 0.2 mas. As for INPOP ephemerides, the 0.5 mas accuracy
of the tie was reported for an older version INPOP13~\cite{INPOP13}. The LLR data is not used for that tie;
however, the lunar solution in DE430~\cite{DE430TR,williams13} contains four corrections
to IAU1980 nutation model, namely:
\begin{itemize}
\item X-axis rotation at J2000 and its rate, ``obliquity rate''
\item Y-axis rotation at J2000 and its rate, ``luni-solar precession''
  \end{itemize}
In DE430, those corrections are not used outside the lunar solution,
and particularly, are not used for ephemeris-ICRF frame tie determination.

Similar four corrections (but to IAU2000 nutation model rather than to IAU1980)
were obtained in~\cite{Hofmann2018Contributions}. The corrections were treated
as \textit{the} ``tie between the dynamical ephemeris frame to the kinematic celestial
frame''; no spacecraft VLBI data were considered.

\begin{figure}
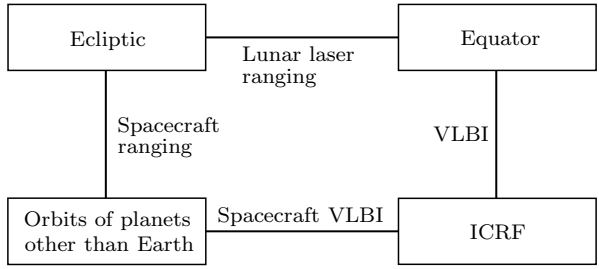

  \input ties.tex
  \caption{Different ties realised with different techniques.}
  \label{fig:ties}
  \end{figure}

In this work, the two methods of obtaining BCRF--ICRF tie are applied and their
results are compared. Fig.~\ref{fig:ties} shows the (seemingly) redundant scheme
to tie planetary orbits to ICRF. The direct ``orbits--ICRF'' connector is what
is normally used in ephemerides. At the same time, the orbits of planets are
tied to the ecliptic in the ephemeris, via spacecraft ranging (most precise
technique), and also differenced range observations of spacecraft and optical observations
of planets and their satellites. The ecliptic is connected to the equator via
LLR. The equator, in turn, is tied to the ICRF via VLBI observations of quasars.

\subsection{Setting of experiment}

Planetary solution (see Sec.~\ref{sec:main-solution}) was obtained and its frame
tied to the ICRF using spacecraft VLBI observations. Then, the lunar solution
was re-obtained, with eight more parameters from~(\ref{eq:t2c}): $\Delta X,
\Delta Y, \dot{X}, \dot{Y}, C^{\Omega}_X$, $S^{\Omega}_X$, $C^{\Omega}_Y$ and $S^{\Omega}_Y$.
(The $\Delta Z$ rotation can not be determined as
long as the positions of the stations are determined in the solution.) These
parameters, per se, have no relation to planets; we call them the tie between
the dynamical Earth--Moon system and the ICRF.  The procedure is repeated four
times: for C04 and ``finals'' EOP series, and for DE430 and IERS models of tidal
variations of geopotential.

\subsection{Results and discussion}

Table~\ref{tbl:icrf} shows the determined rotations, their trends and
18.6-year amplitudes. It is visible that the values
vary greatly across the four solutions. In particular,
the strong $\dot{X}$ trend is detected in all solutions but one. The strong
$\dot{Y}$ trend is detected in two solutions. There is not one solution
where both trends are low. That fact alone makes the determined constant
rotations irrelevant; the trends must be explained first.
The same applies to the determined amplitudes. It is
visible that they are all big with the IERS tidal model;
also, neither of the two solutions based on the DE430 tidal model
has low values of all the four amplitudes.

\begin{table}[h]
  \caption{Ephemeris-ICRF rotation angles, their rates and 18.6 year amplitudes,
    determined from LLR.
     The four columns of numbers are the combinations of two geopotential variations
     models (DE430 and IERS2010) and two IERS EOP series (C04 and ``finals'').
  The given errors are $1\sigma$.} 
  \begin{tabular}{ | l | r | r | r | r | }
    \hline
    \multicolumn{1}{|c|}{} & \multicolumn{2}{c|}{DE430 tides} & \multicolumn{2}{c|}{IERS2010 tides} \\
    \cline{2-5} 
    & \multicolumn{1}{c|}{C04} & \multicolumn{1}{c|}{finals} &
    \multicolumn{1}{c|}{C04} & \multicolumn{1}{c|}{finals} \\ \hline
    $\Delta X_0$, mas & $0.375\pm 0.049$ & $-0.050\pm 0.046$ & $0.221\pm 0.055$& $-0.184\pm 0.052$ \\  \hline
    $\Delta Y_0$, mas & $0.011\pm 0.026$ & $0.052\pm 0.024$ & $0.287\pm 0.029$ & $0.279\pm 0.027$ \\ \hline
    $\dot{X}$, $\mu$as/year & $-30.1\pm 4.1$ & $-13.5\pm 3.8$ & $-14.1\pm 4.6$ & $2.1\pm 4.3$ \\ \hline
    $\dot{Y}$, $\mu$as/year & $10.9\pm 2.2$ & $12.4\pm 2.0$ & $-14.4\pm 2.9$ & $-14.1 \pm 2.3$ \\ \hline
    $C^{\Omega}_X$, mas & $-0.026\pm 0.039$ & $-0.291\pm 0.037$ & $-0.620\pm 0.042$ & $-0.663\pm 0.040$\\ \hline
    $S^{\Omega}_X$, mas & $-0.228\pm 0.034$ & $-0.260\pm 0.031$ & $0.465\pm 0.037$ &  $0.457\pm 0.035$\\ \hline
    $C^{\Omega}_Y$, mas & $0.102\pm 0.026$ & $0.018\pm 0.025$ & $0.505\pm 0.030$ &  $0.414\pm 0.028$\\ \hline
    $S^{\Omega}_Y$, mas & $-0.033\pm 0.036$ & $0.026\pm 0.033$ & $0.329\pm 0.039$ &  $0.388\pm 0.036$\\ \hline
  \end{tabular}
  \label{tbl:icrf}
\end{table}

To further study the temporal behavior of the ICRF tie, four other lunar
solutions were obtained---without the determined trends or amplitudes, but rather with seven
pairs of $(\Delta X, \Delta Y)$, each affecting a seven-year timespan of LLR
observations.  In total, the ``piecewise tie'' to ICRF was calculated for
the timespan of 49 years (from 1 Aug 1970 to 1 Aug 2019).

\begin{figure}
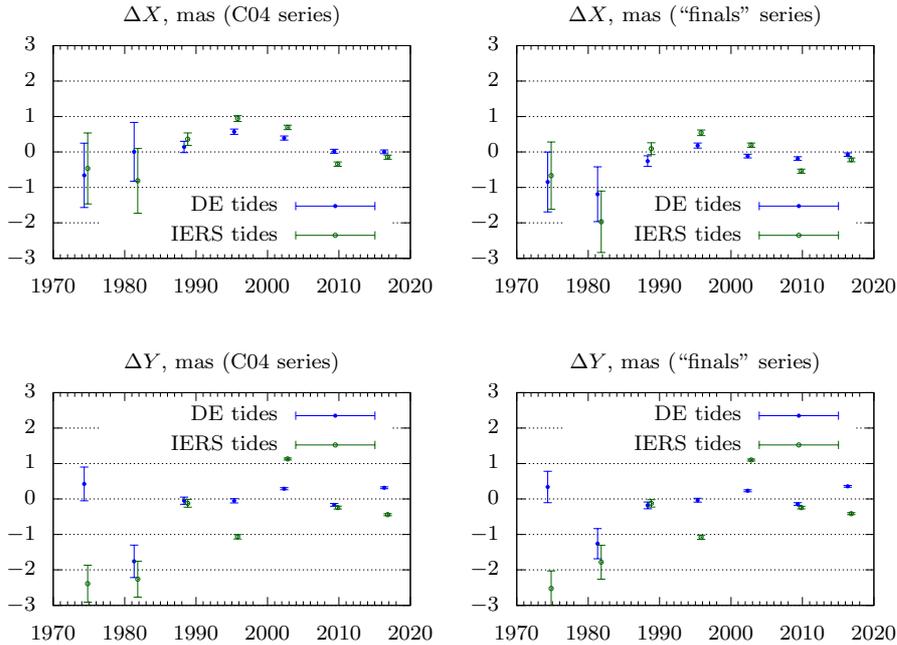

  \input icrf-x-c04.tex
  \input icrf-x-finals.tex
  
  \input icrf-y-c04.tex
  \input icrf-y-finals.tex
  \caption{Piecewise ties for dynamical Earth--Moon system to ICRF. $\Delta X$ and $\Delta Y$
    corrections for different combinations of EOP series (C04/finals) and geopotential
    variations model (DE430/IERS2010) are shown. Each point represents
    a seven-year timespan (starting and ending on 1 May); IERS tides points have been
    artificially moved to the right for clarity.}
  \label{fig:icrf}
\end{figure}

Figure~\ref{fig:icrf} shows the piecewise ICRF tie $\Delta X$ and $\Delta Y$
corrections. The baseline of the plots corresponds to the lunar-planetary ephemeris
oriented to ICRF via $\Delta$DOR observations (see Figs.~\ref{fig:aldel-venus}--\ref{fig:vlba-mars}).
It is clear that there are systematic differences between the ties
depending on the used tidal model, EOP series, and time. Until those issues are
resolved, in the requirement of sub-mas accuracy, the LLR observations are of no
help for the BCRF--ICRF tie. However, the obtained result is interesting by
itself because it indicates deeper problems with the assumptions that were made
on Fig.~\ref{fig:ties}.

One problem is the EOP series. Table~\ref{tbl:icrf} and Fig.~\ref{fig:icrf}
clearly show that the tie between the Earth--Moon system and the ICRF suffers
from absence of sub-mas agreement in the celestial pole trend, let alone its
position, between C04 and ``finals'' series. The disagreement in celestial pole
series has been studied before~\cite{Malkin2012,Malkin2014Nutation,Malkin2017}.
For instance, \cite[Table 3]{Malkin2012} shows $dX$ slope for C04 and ``finals''
series at $15.1\pm1.8$  $\mu$as/year and $1.1\pm1.6$ $\mu$as/year, respectively.  So
the difference is outside the error margin.  For $dY$ slope, the values are
$59.6\pm1.8$ and $50.9\pm1.7$ $\mu$as/year. The difference between those slopes
is not as big as with $dX$ slopes, but there is another problem: the values
are bigger than any $dY$ slope obtained from the single VLBI series constituting
the combined series.

The second problem is the lunar model. Different model values of Earth's $J_2$
and different models of tidal variations of that value can cause different
rotational behavior of the Earth--Moon system. The difference between DE430 model
and IERS2010 model is clearly visible in Fig.~\ref{fig:icrf} and Table~\ref{tbl:icrf}:
for instance, $\dot{Y}$ has different signs depending on the model;
also, 18.6-year amplitudes are always strongly detected with the IERS2010 model,
while some of them are not detected with the DE430 model.

The assumptions and comments on their validity are summarized in
Table~\ref{tbl:assumptions}.  With present state of lunar model, and with
present state of celestial pole EOP series, one can not rely on either of those
things to verify the other. Further development of methods for EOP solutions,
and further development of Earth--Moon dynamical model are needed---whichever
happens first, will help to solve problems with the other.

\begin{table}
  \caption{Assumptions about models and frames, and comments on validity on those assumptions on sub-mas level.}
  \begin{tabular}{|p{5cm}|p{6cm}|}
    \hline
    \multicolumn{1}{|c|}{\textbf{Assumption}} &
    \multicolumn{1}{|c|}{\textbf{Validity}}  \\ \hline
    ICRF forms an inertial frame &
    True (apart from the galactic aberration which is estimated at 5--10 $\mu$as/year) \\ \hline
    Dynamic frame of planets is inertial & True (mathematically) \\ \hline
    Dynamic model of planets accounts for
    all natural phenomena that can have a rotational effect on the Solar system
    detectable by present observations &
    Probably true. One problem in that regard may come from inaccurate value of Sun's $J_2$,
    however it is well determined from MESSENGER observations. Also, the planetary solution
    does not detect rotation from spacecraft VLBI observations at the level of few tens $\mu$as/year.  \\ \hline
    Celestial pole coordinates are known from VLBI observations & False, as shown in \cite{Malkin2012,Malkin2014Nutation,Malkin2017} \\ \hline
    Dynamical frame of the Earth--Moon system is inertial & Unknown, because of the involved non-dynamical models
    of Earth gravitational potential and Earth rotation with EOP \\ \hline
    Dynamical model of the Earth--Moon system accounts for all natural phenomena that can have a rotational effect on the real Earth--Moon system detectable by present observations & Probably false, since different models of geopotential already produce different rotational effects, as shown in Table~\ref{tbl:icrf} and Fig.~\ref{fig:icrf}. \\ \hline
   \end{tabular}
  \label{tbl:assumptions}
  \end{table}

\section*{Conclusion}\label{sec:conclusion}

\begin{itemize}
\item OCA observatory (former CERGA) in Grasse, France continues to provide
  large amounts of LLR green and infrared data. Matera observatory in Italy
  provides infrequent LLR data. Wettzell observatory started to provide LLR
  infrared data of a very good quality.
\item LLR was useful for building EOP series in the past. Modern LLR, too, is able to
  detect inaccuracies in modern EOP series, with sub-mas accuracy. Probably the
  LLR data can benefit combined EOP solutions; currently it is used in KEOF,
  which does not include celestial pole.
\item LLR is capable of detecting two out of three lunar orientation parameters
  (LOPs) with accuracy of few mas. However, on the present data, no
  statistically significant daily deviations of LOPs from the lunar rotational
  model were detected. The lunar model research should continue, though, in the
  area of long-term variations.
\item The tie between ephemeris frame and ICRF, calculated from spacecraft VLBI
  ($\Delta$DOR) data, is confirmed with the latest data with the accuracy of
  0.18 mas (3$\sigma$).
\item IERS Bulleting A weekly EOP series produce generally better fits of LLR solutions
  than IERS C04.
\item LLR is potentially capable of tying the Earth--Moon system
  to ICRF (and hence, the whole ephemeris frame to ICRF) with accuracy
  comparable to that of $\Delta$DOR-based tie; however, one obstacle is
  the location of the celestial pole in EOP series: it is not accurate enough to use
  the equator--ICRF link to tie the ecliptic to ICRF via LLR observations and
  their ecliptic--equator link.
\item More research is needed in the area of Earth--Moon dynamical system,
  and particularly in the model of geopotential which affects the lunar
  orbital motion. Two available models (IERS2010 and DE430) produce different
  rotational rates of the Earth--Moon system in the celestial frame, which
  is detectable by LLR observations. The research of the Earth--Moon dynamics
  will be facilitated by an improvement in present celestial pole series (or vice versa).
  \end{itemize}

\begin{acknowledgements}
  Author would like to thank his colleagues: Elena Pitjeva for her
  continuous support and particular help with the spacecraft VLBI topic, and
  Sergey Kurdubov for much helpful advice throughout this work.

  This work would not have been possible without the effort of personnel at
  observatories doing lunar laser ranging, particularly Tom Murphy in Apache Point,
  Etienne Samain and Jean-Marie Torre in Grasse, Giuseppe Bianco in Matera,
  Karl Ulrich Schreiber in Wettzell, Peter Shelus in McDonald, Walter Bonsack in Haleakala,
  and Andrei Severny in Nauchny.
  
  The POLAC  website was of great help, where Christophe Barache, S\'ebastien Bouquillon,
  Teddy Carlucci, and Gerard Francou carefully collected LLR observations from
  different sources.

  Planetary observations, including spacecraft VLBI which is essential for part of this
  work, were collected at the NASA SSD webpage and are supported by William Folkner,
  and at the Geoazur website supported by Agn{\`e}s Fienga.

  Anonymous reviewers provided useful advice which allowed to significantly
  improve the paper.

  \bigskip

  \noindent \textbf{Author contributions}\hskip .7em D. P. designed and performed the research
  and wrote the manuscript. 

  \bigskip

  \noindent \textbf{Data availability}\hskip .7em All the observational data
  used in this work comes from publicly available sources listed in section 2.
  The EOP series are publicly available at the IERS website \url{http://iers.org}.
\end{acknowledgements}

\bibliographystyle{spmpsci}
\bibliography{llr-frames} 

\end{document}

%% file: vlbi-venus.tex
\begingroup
  \makeatletter
  \providecommand\color[2][]{%
    \GenericError{(gnuplot) \space\space\space\@spaces}{%
      Package color not loaded in conjunction with
      terminal option `colourtext'%
    }{See the gnuplot documentation for explanation.%
    }{Either use 'blacktext' in gnuplot or load the package
      color.sty in LaTeX.}%
    \renewcommand\color[2][]{}%
  }%
  \providecommand\includegraphics[2][]{%
    \GenericError{(gnuplot) \space\space\space\@spaces}{%
      Package graphicx or graphics not loaded%
    }{See the gnuplot documentation for explanation.%
    }{The gnuplot epslatex terminal needs graphicx.sty or graphics.sty.}%
    \renewcommand\includegraphics[2][]{}%
  }%
  \providecommand\rotatebox[2]{#2}%
  \@ifundefined{ifGPcolor}{%
    \newif\ifGPcolor
    \GPcolorfalse
  }{}%
  \@ifundefined{ifGPblacktext}{%
    \newif\ifGPblacktext
    \GPblacktexttrue
  }{}%
  \let\gplgaddtomacro\g@addto@macro
  \gdef\gplbacktext{}%
  \gdef\gplfronttext{}%
  \makeatother
  \ifGPblacktext
    \def\colorrgb#1{}%
    \def\colorgray#1{}%
  \else
    \ifGPcolor
      \def\colorrgb#1{\color[rgb]{#1}}%
      \def\colorgray#1{\color[gray]{#1}}%
      \expandafter\def\csname LTw\endcsname{\color{white}}%
      \expandafter\def\csname LTb\endcsname{\color{black}}%
      \expandafter\def\csname LTa\endcsname{\color{black}}%
      \expandafter\def\csname LT0\endcsname{\color[rgb]{1,0,0}}%
      \expandafter\def\csname LT1\endcsname{\color[rgb]{0,1,0}}%
      \expandafter\def\csname LT2\endcsname{\color[rgb]{0,0,1}}%
      \expandafter\def\csname LT3\endcsname{\color[rgb]{1,0,1}}%
      \expandafter\def\csname LT4\endcsname{\color[rgb]{0,1,1}}%
      \expandafter\def\csname LT5\endcsname{\color[rgb]{1,1,0}}%
      \expandafter\def\csname LT6\endcsname{\color[rgb]{0,0,0}}%
      \expandafter\def\csname LT7\endcsname{\color[rgb]{1,0.3,0}}%
      \expandafter\def\csname LT8\endcsname{\color[rgb]{0.5,0.5,0.5}}%
    \else
      \def\colorrgb#1{\color{black}}%
      \def\colorgray#1{\color[gray]{#1}}%
      \expandafter\def\csname LTw\endcsname{\color{white}}%
      \expandafter\def\csname LTb\endcsname{\color{black}}%
      \expandafter\def\csname LTa\endcsname{\color{black}}%
      \expandafter\def\csname LT0\endcsname{\color{black}}%
      \expandafter\def\csname LT1\endcsname{\color{black}}%
      \expandafter\def\csname LT2\endcsname{\color{black}}%
      \expandafter\def\csname LT3\endcsname{\color{black}}%
      \expandafter\def\csname LT4\endcsname{\color{black}}%
      \expandafter\def\csname LT5\endcsname{\color{black}}%
      \expandafter\def\csname LT6\endcsname{\color{black}}%
      \expandafter\def\csname LT7\endcsname{\color{black}}%
      \expandafter\def\csname LT8\endcsname{\color{black}}%
    \fi
  \fi
    \setlength{\unitlength}{0.0500bp}%
    \ifx\gptboxheight\undefined%
      \newlength{\gptboxheight}%
      \newlength{\gptboxwidth}%
      \newsavebox{\gptboxtext}%
    \fi%
    \setlength{\fboxrule}{0.5pt}%
    \setlength{\fboxsep}{1pt}%
\begin{picture}(5040.00,2304.00)%
    \gplgaddtomacro\gplbacktext{%
      \csname LTb\endcsname%
      \put(616,440){\makebox(0,0)[r]{\strut{}$-10$}}%
      \csname LTb\endcsname%
      \put(616,604){\makebox(0,0)[r]{\strut{}$-8$}}%
      \csname LTb\endcsname%
      \put(616,769){\makebox(0,0)[r]{\strut{}$-6$}}%
      \csname LTb\endcsname%
      \put(616,933){\makebox(0,0)[r]{\strut{}$-4$}}%
      \csname LTb\endcsname%
      \put(616,1097){\makebox(0,0)[r]{\strut{}$-2$}}%
      \csname LTb\endcsname%
      \put(616,1262){\makebox(0,0)[r]{\strut{}$0$}}%
      \csname LTb\endcsname%
      \put(616,1426){\makebox(0,0)[r]{\strut{}$2$}}%
      \csname LTb\endcsname%
      \put(616,1590){\makebox(0,0)[r]{\strut{}$4$}}%
      \csname LTb\endcsname%
      \put(616,1754){\makebox(0,0)[r]{\strut{}$6$}}%
      \csname LTb\endcsname%
      \put(616,1919){\makebox(0,0)[r]{\strut{}$8$}}%
      \csname LTb\endcsname%
      \put(616,2083){\makebox(0,0)[r]{\strut{}$10$}}%
      \put(2696,220){\makebox(0,0){\strut{}2006}}%
      \put(3182,220){\makebox(0,0){\strut{}2008}}%
      \put(3669,220){\makebox(0,0){\strut{}2010}}%
      \put(4156,220){\makebox(0,0){\strut{}2012}}%
      \put(4643,220){\makebox(0,0){\strut{}2014}}%
      \put(748,220){\makebox(0,0){\strut{}1990}}%
      \put(1235,220){\makebox(0,0){\strut{}1992}}%
      \put(1722,220){\makebox(0,0){\strut{}1994}}%
      \put(2208,220){\makebox(0,0){\strut{}1996}}%
    }%
    \gplgaddtomacro\gplfronttext{%
      \csname LTb\endcsname%
      \put(176,1261){\rotatebox{-270}{\makebox(0,0){\strut{}$O - C$, mas}}}%
      \put(2695,2303){\makebox(0,0){\strut{}Venus, single-baseline VLBI}}%
    }%
    \gplbacktext
    \put(0,0){\includegraphics{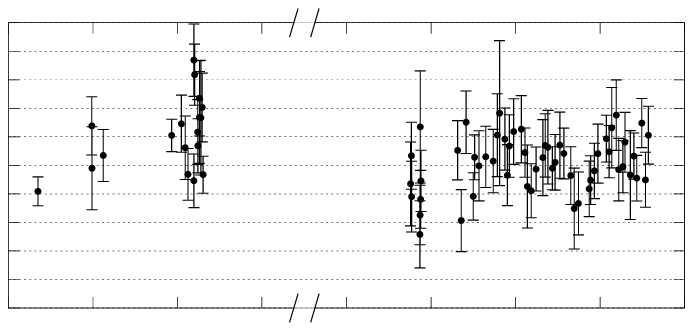}}%
    \gplfronttext
  \end{picture}%
\endgroup

%% file: vlbi-mars.tex
\begingroup
  \makeatletter
  \providecommand\color[2][]{%
    \GenericError{(gnuplot) \space\space\space\@spaces}{%
      Package color not loaded in conjunction with
      terminal option `colourtext'%
    }{See the gnuplot documentation for explanation.%
    }{Either use 'blacktext' in gnuplot or load the package
      color.sty in LaTeX.}%
    \renewcommand\color[2][]{}%
  }%
  \providecommand\includegraphics[2][]{%
    \GenericError{(gnuplot) \space\space\space\@spaces}{%
      Package graphicx or graphics not loaded%
    }{See the gnuplot documentation for explanation.%
    }{The gnuplot epslatex terminal needs graphicx.sty or graphics.sty.}%
    \renewcommand\includegraphics[2][]{}%
  }%
  \providecommand\rotatebox[2]{#2}%
  \@ifundefined{ifGPcolor}{%
    \newif\ifGPcolor
    \GPcolorfalse
  }{}%
  \@ifundefined{ifGPblacktext}{%
    \newif\ifGPblacktext
    \GPblacktexttrue
  }{}%
  \let\gplgaddtomacro\g@addto@macro
  \gdef\gplbacktext{}%
  \gdef\gplfronttext{}%
  \makeatother
  \ifGPblacktext
    \def\colorrgb#1{}%
    \def\colorgray#1{}%
  \else
    \ifGPcolor
      \def\colorrgb#1{\color[rgb]{#1}}%
      \def\colorgray#1{\color[gray]{#1}}%
      \expandafter\def\csname LTw\endcsname{\color{white}}%
      \expandafter\def\csname LTb\endcsname{\color{black}}%
      \expandafter\def\csname LTa\endcsname{\color{black}}%
      \expandafter\def\csname LT0\endcsname{\color[rgb]{1,0,0}}%
      \expandafter\def\csname LT1\endcsname{\color[rgb]{0,1,0}}%
      \expandafter\def\csname LT2\endcsname{\color[rgb]{0,0,1}}%
      \expandafter\def\csname LT3\endcsname{\color[rgb]{1,0,1}}%
      \expandafter\def\csname LT4\endcsname{\color[rgb]{0,1,1}}%
      \expandafter\def\csname LT5\endcsname{\color[rgb]{1,1,0}}%
      \expandafter\def\csname LT6\endcsname{\color[rgb]{0,0,0}}%
      \expandafter\def\csname LT7\endcsname{\color[rgb]{1,0.3,0}}%
      \expandafter\def\csname LT8\endcsname{\color[rgb]{0.5,0.5,0.5}}%
    \else
      \def\colorrgb#1{\color{black}}%
      \def\colorgray#1{\color[gray]{#1}}%
      \expandafter\def\csname LTw\endcsname{\color{white}}%
      \expandafter\def\csname LTb\endcsname{\color{black}}%
      \expandafter\def\csname LTa\endcsname{\color{black}}%
      \expandafter\def\csname LT0\endcsname{\color{black}}%
      \expandafter\def\csname LT1\endcsname{\color{black}}%
      \expandafter\def\csname LT2\endcsname{\color{black}}%
      \expandafter\def\csname LT3\endcsname{\color{black}}%
      \expandafter\def\csname LT4\endcsname{\color{black}}%
      \expandafter\def\csname LT5\endcsname{\color{black}}%
      \expandafter\def\csname LT6\endcsname{\color{black}}%
      \expandafter\def\csname LT7\endcsname{\color{black}}%
      \expandafter\def\csname LT8\endcsname{\color{black}}%
    \fi
  \fi
    \setlength{\unitlength}{0.0500bp}%
    \ifx\gptboxheight\undefined%
      \newlength{\gptboxheight}%
      \newlength{\gptboxwidth}%
      \newsavebox{\gptboxtext}%
    \fi%
    \setlength{\fboxrule}{0.5pt}%
    \setlength{\fboxsep}{1pt}%
\begin{picture}(4030.00,2304.00)%
    \gplgaddtomacro\gplbacktext{%
      \csname LTb\endcsname%
      \put(748,440){\makebox(0,0)[r]{\strut{}$-2.5$}}%
      \csname LTb\endcsname%
      \put(748,604){\makebox(0,0)[r]{\strut{}$-2$}}%
      \csname LTb\endcsname%
      \put(748,769){\makebox(0,0)[r]{\strut{}$-1.5$}}%
      \csname LTb\endcsname%
      \put(748,933){\makebox(0,0)[r]{\strut{}$-1$}}%
      \csname LTb\endcsname%
      \put(748,1097){\makebox(0,0)[r]{\strut{}$-0.5$}}%
      \csname LTb\endcsname%
      \put(748,1262){\makebox(0,0)[r]{\strut{}$0$}}%
      \csname LTb\endcsname%
      \put(748,1426){\makebox(0,0)[r]{\strut{}$0.5$}}%
      \csname LTb\endcsname%
      \put(748,1590){\makebox(0,0)[r]{\strut{}$1$}}%
      \csname LTb\endcsname%
      \put(748,1754){\makebox(0,0)[r]{\strut{}$1.5$}}%
      \csname LTb\endcsname%
      \put(748,1919){\makebox(0,0)[r]{\strut{}$2$}}%
      \csname LTb\endcsname%
      \put(748,2083){\makebox(0,0)[r]{\strut{}$2.5$}}%
      \put(880,220){\makebox(0,0){\strut{}2000}}%
      \put(1273,220){\makebox(0,0){\strut{}2002}}%
      \put(1666,220){\makebox(0,0){\strut{}2004}}%
      \put(2060,220){\makebox(0,0){\strut{}2006}}%
      \put(2453,220){\makebox(0,0){\strut{}2008}}%
      \put(2846,220){\makebox(0,0){\strut{}2010}}%
      \put(3239,220){\makebox(0,0){\strut{}2012}}%
      \put(3632,220){\makebox(0,0){\strut{}2014}}%
    }%
    \gplgaddtomacro\gplfronttext{%
      \csname LTb\endcsname%
      \put(176,1261){\rotatebox{-270}{\makebox(0,0){\strut{}$O - C$, mas}}}%
      \put(2256,2303){\makebox(0,0){\strut{}Mars, single-baseline VLBI}}%
    }%
    \gplbacktext
    \put(0,0){\includegraphics{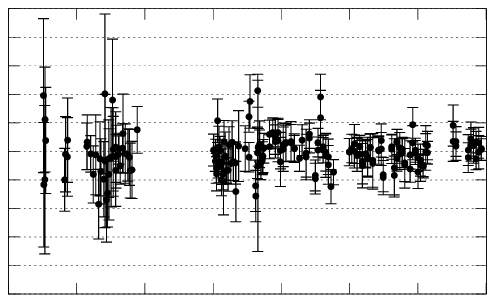}}%
    \gplfronttext
  \end{picture}%
\endgroup

%% file: vlbi-jupiter.tex
\begingroup
  \makeatletter
  \providecommand\color[2][]{%
    \GenericError{(gnuplot) \space\space\space\@spaces}{%
      Package color not loaded in conjunction with
      terminal option `colourtext'%
    }{See the gnuplot documentation for explanation.%
    }{Either use 'blacktext' in gnuplot or load the package
      color.sty in LaTeX.}%
    \renewcommand\color[2][]{}%
  }%
  \providecommand\includegraphics[2][]{%
    \GenericError{(gnuplot) \space\space\space\@spaces}{%
      Package graphicx or graphics not loaded%
    }{See the gnuplot documentation for explanation.%
    }{The gnuplot epslatex terminal needs graphicx.sty or graphics.sty.}%
    \renewcommand\includegraphics[2][]{}%
  }%
  \providecommand\rotatebox[2]{#2}%
  \@ifundefined{ifGPcolor}{%
    \newif\ifGPcolor
    \GPcolorfalse
  }{}%
  \@ifundefined{ifGPblacktext}{%
    \newif\ifGPblacktext
    \GPblacktexttrue
  }{}%
  \let\gplgaddtomacro\g@addto@macro
  \gdef\gplbacktext{}%
  \gdef\gplfronttext{}%
  \makeatother
  \ifGPblacktext
    \def\colorrgb#1{}%
    \def\colorgray#1{}%
  \else
    \ifGPcolor
      \def\colorrgb#1{\color[rgb]{#1}}%
      \def\colorgray#1{\color[gray]{#1}}%
      \expandafter\def\csname LTw\endcsname{\color{white}}%
      \expandafter\def\csname LTb\endcsname{\color{black}}%
      \expandafter\def\csname LTa\endcsname{\color{black}}%
      \expandafter\def\csname LT0\endcsname{\color[rgb]{1,0,0}}%
      \expandafter\def\csname LT1\endcsname{\color[rgb]{0,1,0}}%
      \expandafter\def\csname LT2\endcsname{\color[rgb]{0,0,1}}%
      \expandafter\def\csname LT3\endcsname{\color[rgb]{1,0,1}}%
      \expandafter\def\csname LT4\endcsname{\color[rgb]{0,1,1}}%
      \expandafter\def\csname LT5\endcsname{\color[rgb]{1,1,0}}%
      \expandafter\def\csname LT6\endcsname{\color[rgb]{0,0,0}}%
      \expandafter\def\csname LT7\endcsname{\color[rgb]{1,0.3,0}}%
      \expandafter\def\csname LT8\endcsname{\color[rgb]{0.5,0.5,0.5}}%
    \else
      \def\colorrgb#1{\color{black}}%
      \def\colorgray#1{\color[gray]{#1}}%
      \expandafter\def\csname LTw\endcsname{\color{white}}%
      \expandafter\def\csname LTb\endcsname{\color{black}}%
      \expandafter\def\csname LTa\endcsname{\color{black}}%
      \expandafter\def\csname LT0\endcsname{\color{black}}%
      \expandafter\def\csname LT1\endcsname{\color{black}}%
      \expandafter\def\csname LT2\endcsname{\color{black}}%
      \expandafter\def\csname LT3\endcsname{\color{black}}%
      \expandafter\def\csname LT4\endcsname{\color{black}}%
      \expandafter\def\csname LT5\endcsname{\color{black}}%
      \expandafter\def\csname LT6\endcsname{\color{black}}%
      \expandafter\def\csname LT7\endcsname{\color{black}}%
      \expandafter\def\csname LT8\endcsname{\color{black}}%
    \fi
  \fi
    \setlength{\unitlength}{0.0500bp}%
    \ifx\gptboxheight\undefined%
      \newlength{\gptboxheight}%
      \newlength{\gptboxwidth}%
      \newsavebox{\gptboxtext}%
    \fi%
    \setlength{\fboxrule}{0.5pt}%
    \setlength{\fboxsep}{1pt}%
\begin{picture}(2880.00,2304.00)%
    \gplgaddtomacro\gplbacktext{%
      \csname LTb\endcsname%
      \put(616,440){\makebox(0,0)[r]{\strut{}$-25$}}%
      \csname LTb\endcsname%
      \put(616,604){\makebox(0,0)[r]{\strut{}$-20$}}%
      \csname LTb\endcsname%
      \put(616,769){\makebox(0,0)[r]{\strut{}$-15$}}%
      \csname LTb\endcsname%
      \put(616,933){\makebox(0,0)[r]{\strut{}$-10$}}%
      \csname LTb\endcsname%
      \put(616,1097){\makebox(0,0)[r]{\strut{}$-5$}}%
      \csname LTb\endcsname%
      \put(616,1262){\makebox(0,0)[r]{\strut{}$0$}}%
      \csname LTb\endcsname%
      \put(616,1426){\makebox(0,0)[r]{\strut{}$5$}}%
      \csname LTb\endcsname%
      \put(616,1590){\makebox(0,0)[r]{\strut{}$10$}}%
      \csname LTb\endcsname%
      \put(616,1754){\makebox(0,0)[r]{\strut{}$15$}}%
      \csname LTb\endcsname%
      \put(616,1919){\makebox(0,0)[r]{\strut{}$20$}}%
      \csname LTb\endcsname%
      \put(616,2083){\makebox(0,0)[r]{\strut{}$25$}}%
      \put(748,220){\makebox(0,0){\strut{}1996}}%
      \put(1616,220){\makebox(0,0){\strut{}1997}}%
      \put(2481,220){\makebox(0,0){\strut{}1998}}%
    }%
    \gplgaddtomacro\gplfronttext{%
      \csname LTb\endcsname%
      \put(176,1261){\rotatebox{-270}{\makebox(0,0){\strut{}$O - C$, mas}}}%
      \put(1615,2303){\makebox(0,0){\strut{}Jupiter, single-baseline VLBI}}%
    }%
    \gplbacktext
    \put(0,0){\includegraphics{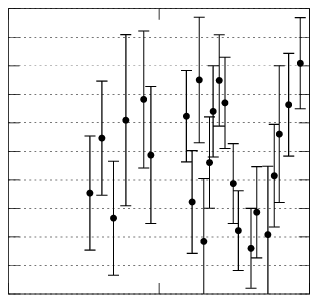}}%
    \gplfronttext
  \end{picture}%
\endgroup

%% file: vlba-saturn-alpha.tex
\begingroup
  \makeatletter
  \providecommand\color[2][]{%
    \GenericError{(gnuplot) \space\space\space\@spaces}{%
      Package color not loaded in conjunction with
      terminal option `colourtext'%
    }{See the gnuplot documentation for explanation.%
    }{Either use 'blacktext' in gnuplot or load the package
      color.sty in LaTeX.}%
    \renewcommand\color[2][]{}%
  }%
  \providecommand\includegraphics[2][]{%
    \GenericError{(gnuplot) \space\space\space\@spaces}{%
      Package graphicx or graphics not loaded%
    }{See the gnuplot documentation for explanation.%
    }{The gnuplot epslatex terminal needs graphicx.sty or graphics.sty.}%
    \renewcommand\includegraphics[2][]{}%
  }%
  \providecommand\rotatebox[2]{#2}%
  \@ifundefined{ifGPcolor}{%
    \newif\ifGPcolor
    \GPcolorfalse
  }{}%
  \@ifundefined{ifGPblacktext}{%
    \newif\ifGPblacktext
    \GPblacktexttrue
  }{}%
  \let\gplgaddtomacro\g@addto@macro
  \gdef\gplbacktext{}%
  \gdef\gplfronttext{}%
  \makeatother
  \ifGPblacktext
    \def\colorrgb#1{}%
    \def\colorgray#1{}%
  \else
    \ifGPcolor
      \def\colorrgb#1{\color[rgb]{#1}}%
      \def\colorgray#1{\color[gray]{#1}}%
      \expandafter\def\csname LTw\endcsname{\color{white}}%
      \expandafter\def\csname LTb\endcsname{\color{black}}%
      \expandafter\def\csname LTa\endcsname{\color{black}}%
      \expandafter\def\csname LT0\endcsname{\color[rgb]{1,0,0}}%
      \expandafter\def\csname LT1\endcsname{\color[rgb]{0,1,0}}%
      \expandafter\def\csname LT2\endcsname{\color[rgb]{0,0,1}}%
      \expandafter\def\csname LT3\endcsname{\color[rgb]{1,0,1}}%
      \expandafter\def\csname LT4\endcsname{\color[rgb]{0,1,1}}%
      \expandafter\def\csname LT5\endcsname{\color[rgb]{1,1,0}}%
      \expandafter\def\csname LT6\endcsname{\color[rgb]{0,0,0}}%
      \expandafter\def\csname LT7\endcsname{\color[rgb]{1,0.3,0}}%
      \expandafter\def\csname LT8\endcsname{\color[rgb]{0.5,0.5,0.5}}%
    \else
      \def\colorrgb#1{\color{black}}%
      \def\colorgray#1{\color[gray]{#1}}%
      \expandafter\def\csname LTw\endcsname{\color{white}}%
      \expandafter\def\csname LTb\endcsname{\color{black}}%
      \expandafter\def\csname LTa\endcsname{\color{black}}%
      \expandafter\def\csname LT0\endcsname{\color{black}}%
      \expandafter\def\csname LT1\endcsname{\color{black}}%
      \expandafter\def\csname LT2\endcsname{\color{black}}%
      \expandafter\def\csname LT3\endcsname{\color{black}}%
      \expandafter\def\csname LT4\endcsname{\color{black}}%
      \expandafter\def\csname LT5\endcsname{\color{black}}%
      \expandafter\def\csname LT6\endcsname{\color{black}}%
      \expandafter\def\csname LT7\endcsname{\color{black}}%
      \expandafter\def\csname LT8\endcsname{\color{black}}%
    \fi
  \fi
    \setlength{\unitlength}{0.0500bp}%
    \ifx\gptboxheight\undefined%
      \newlength{\gptboxheight}%
      \newlength{\gptboxwidth}%
      \newsavebox{\gptboxtext}%
    \fi%
    \setlength{\fboxrule}{0.5pt}%
    \setlength{\fboxsep}{1pt}%
\begin{picture}(3528.00,2160.00)%
    \gplgaddtomacro\gplbacktext{%
      \csname LTb\endcsname%
      \put(726,440){\makebox(0,0)[r]{\strut{}$-2$}}%
      \csname LTb\endcsname%
      \put(726,627){\makebox(0,0)[r]{\strut{}$-1.5$}}%
      \csname LTb\endcsname%
      \put(726,815){\makebox(0,0)[r]{\strut{}$-1$}}%
      \csname LTb\endcsname%
      \put(726,1002){\makebox(0,0)[r]{\strut{}$-0.5$}}%
      \csname LTb\endcsname%
      \put(726,1190){\makebox(0,0)[r]{\strut{}$0$}}%
      \csname LTb\endcsname%
      \put(726,1377){\makebox(0,0)[r]{\strut{}$0.5$}}%
      \csname LTb\endcsname%
      \put(726,1564){\makebox(0,0)[r]{\strut{}$1$}}%
      \csname LTb\endcsname%
      \put(726,1752){\makebox(0,0)[r]{\strut{}$1.5$}}%
      \csname LTb\endcsname%
      \put(726,1939){\makebox(0,0)[r]{\strut{}$2$}}%
      \put(858,220){\makebox(0,0){\strut{}2004}}%
      \put(1271,220){\makebox(0,0){\strut{}2006}}%
      \put(1684,220){\makebox(0,0){\strut{}2008}}%
      \put(2098,220){\makebox(0,0){\strut{}2010}}%
      \put(2511,220){\makebox(0,0){\strut{}2012}}%
      \put(2924,220){\makebox(0,0){\strut{}2014}}%
    }%
    \gplgaddtomacro\gplfronttext{%
      \csname LTb\endcsname%
      \put(220,1189){\rotatebox{-270}{\makebox(0,0){\strut{}$O - C$, mas}}}%
      \put(1994,2159){\makebox(0,0){\strut{}Saturn, VLBA, $\alpha$}}%
    }%
    \gplbacktext
    \put(0,0){\includegraphics{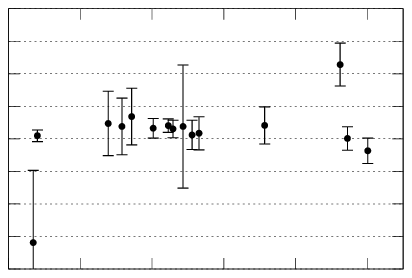}}%
    \gplfronttext
  \end{picture}%
\endgroup

%% file: vlba-saturn-delta.tex
\begingroup
  \makeatletter
  \providecommand\color[2][]{%
    \GenericError{(gnuplot) \space\space\space\@spaces}{%
      Package color not loaded in conjunction with
      terminal option `colourtext'%
    }{See the gnuplot documentation for explanation.%
    }{Either use 'blacktext' in gnuplot or load the package
      color.sty in LaTeX.}%
    \renewcommand\color[2][]{}%
  }%
  \providecommand\includegraphics[2][]{%
    \GenericError{(gnuplot) \space\space\space\@spaces}{%
      Package graphicx or graphics not loaded%
    }{See the gnuplot documentation for explanation.%
    }{The gnuplot epslatex terminal needs graphicx.sty or graphics.sty.}%
    \renewcommand\includegraphics[2][]{}%
  }%
  \providecommand\rotatebox[2]{#2}%
  \@ifundefined{ifGPcolor}{%
    \newif\ifGPcolor
    \GPcolorfalse
  }{}%
  \@ifundefined{ifGPblacktext}{%
    \newif\ifGPblacktext
    \GPblacktexttrue
  }{}%
  \let\gplgaddtomacro\g@addto@macro
  \gdef\gplbacktext{}%
  \gdef\gplfronttext{}%
  \makeatother
  \ifGPblacktext
    \def\colorrgb#1{}%
    \def\colorgray#1{}%
  \else
    \ifGPcolor
      \def\colorrgb#1{\color[rgb]{#1}}%
      \def\colorgray#1{\color[gray]{#1}}%
      \expandafter\def\csname LTw\endcsname{\color{white}}%
      \expandafter\def\csname LTb\endcsname{\color{black}}%
      \expandafter\def\csname LTa\endcsname{\color{black}}%
      \expandafter\def\csname LT0\endcsname{\color[rgb]{1,0,0}}%
      \expandafter\def\csname LT1\endcsname{\color[rgb]{0,1,0}}%
      \expandafter\def\csname LT2\endcsname{\color[rgb]{0,0,1}}%
      \expandafter\def\csname LT3\endcsname{\color[rgb]{1,0,1}}%
      \expandafter\def\csname LT4\endcsname{\color[rgb]{0,1,1}}%
      \expandafter\def\csname LT5\endcsname{\color[rgb]{1,1,0}}%
      \expandafter\def\csname LT6\endcsname{\color[rgb]{0,0,0}}%
      \expandafter\def\csname LT7\endcsname{\color[rgb]{1,0.3,0}}%
      \expandafter\def\csname LT8\endcsname{\color[rgb]{0.5,0.5,0.5}}%
    \else
      \def\colorrgb#1{\color{black}}%
      \def\colorgray#1{\color[gray]{#1}}%
      \expandafter\def\csname LTw\endcsname{\color{white}}%
      \expandafter\def\csname LTb\endcsname{\color{black}}%
      \expandafter\def\csname LTa\endcsname{\color{black}}%
      \expandafter\def\csname LT0\endcsname{\color{black}}%
      \expandafter\def\csname LT1\endcsname{\color{black}}%
      \expandafter\def\csname LT2\endcsname{\color{black}}%
      \expandafter\def\csname LT3\endcsname{\color{black}}%
      \expandafter\def\csname LT4\endcsname{\color{black}}%
      \expandafter\def\csname LT5\endcsname{\color{black}}%
      \expandafter\def\csname LT6\endcsname{\color{black}}%
      \expandafter\def\csname LT7\endcsname{\color{black}}%
      \expandafter\def\csname LT8\endcsname{\color{black}}%
    \fi
  \fi
    \setlength{\unitlength}{0.0500bp}%
    \ifx\gptboxheight\undefined%
      \newlength{\gptboxheight}%
      \newlength{\gptboxwidth}%
      \newsavebox{\gptboxtext}%
    \fi%
    \setlength{\fboxrule}{0.5pt}%
    \setlength{\fboxsep}{1pt}%
\begin{picture}(3528.00,2160.00)%
    \gplgaddtomacro\gplbacktext{%
      \csname LTb\endcsname%
      \put(726,440){\makebox(0,0)[r]{\strut{}$-2$}}%
      \csname LTb\endcsname%
      \put(726,627){\makebox(0,0)[r]{\strut{}$-1.5$}}%
      \csname LTb\endcsname%
      \put(726,815){\makebox(0,0)[r]{\strut{}$-1$}}%
      \csname LTb\endcsname%
      \put(726,1002){\makebox(0,0)[r]{\strut{}$-0.5$}}%
      \csname LTb\endcsname%
      \put(726,1190){\makebox(0,0)[r]{\strut{}$0$}}%
      \csname LTb\endcsname%
      \put(726,1377){\makebox(0,0)[r]{\strut{}$0.5$}}%
      \csname LTb\endcsname%
      \put(726,1564){\makebox(0,0)[r]{\strut{}$1$}}%
      \csname LTb\endcsname%
      \put(726,1752){\makebox(0,0)[r]{\strut{}$1.5$}}%
      \csname LTb\endcsname%
      \put(726,1939){\makebox(0,0)[r]{\strut{}$2$}}%
      \put(858,220){\makebox(0,0){\strut{}2004}}%
      \put(1271,220){\makebox(0,0){\strut{}2006}}%
      \put(1684,220){\makebox(0,0){\strut{}2008}}%
      \put(2098,220){\makebox(0,0){\strut{}2010}}%
      \put(2511,220){\makebox(0,0){\strut{}2012}}%
      \put(2924,220){\makebox(0,0){\strut{}2014}}%
    }%
    \gplgaddtomacro\gplfronttext{%
      \csname LTb\endcsname%
      \put(220,1189){\rotatebox{-270}{\makebox(0,0){\strut{}$O - C$, mas}}}%
      \put(1994,2159){\makebox(0,0){\strut{}Saturn, VLBA, $\delta$}}%
    }%
    \gplbacktext
    \put(0,0){\includegraphics{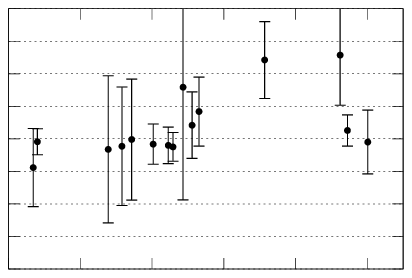}}%
    \gplfronttext
  \end{picture}%
\endgroup

%% file: vlba-mars-alpha.tex
\begingroup
  \makeatletter
  \providecommand\color[2][]{%
    \GenericError{(gnuplot) \space\space\space\@spaces}{%
      Package color not loaded in conjunction with
      terminal option `colourtext'%
    }{See the gnuplot documentation for explanation.%
    }{Either use 'blacktext' in gnuplot or load the package
      color.sty in LaTeX.}%
    \renewcommand\color[2][]{}%
  }%
  \providecommand\includegraphics[2][]{%
    \GenericError{(gnuplot) \space\space\space\@spaces}{%
      Package graphicx or graphics not loaded%
    }{See the gnuplot documentation for explanation.%
    }{The gnuplot epslatex terminal needs graphicx.sty or graphics.sty.}%
    \renewcommand\includegraphics[2][]{}%
  }%
  \providecommand\rotatebox[2]{#2}%
  \@ifundefined{ifGPcolor}{%
    \newif\ifGPcolor
    \GPcolorfalse
  }{}%
  \@ifundefined{ifGPblacktext}{%
    \newif\ifGPblacktext
    \GPblacktexttrue
  }{}%
  \let\gplgaddtomacro\g@addto@macro
  \gdef\gplbacktext{}%
  \gdef\gplfronttext{}%
  \makeatother
  \ifGPblacktext
    \def\colorrgb#1{}%
    \def\colorgray#1{}%
  \else
    \ifGPcolor
      \def\colorrgb#1{\color[rgb]{#1}}%
      \def\colorgray#1{\color[gray]{#1}}%
      \expandafter\def\csname LTw\endcsname{\color{white}}%
      \expandafter\def\csname LTb\endcsname{\color{black}}%
      \expandafter\def\csname LTa\endcsname{\color{black}}%
      \expandafter\def\csname LT0\endcsname{\color[rgb]{1,0,0}}%
      \expandafter\def\csname LT1\endcsname{\color[rgb]{0,1,0}}%
      \expandafter\def\csname LT2\endcsname{\color[rgb]{0,0,1}}%
      \expandafter\def\csname LT3\endcsname{\color[rgb]{1,0,1}}%
      \expandafter\def\csname LT4\endcsname{\color[rgb]{0,1,1}}%
      \expandafter\def\csname LT5\endcsname{\color[rgb]{1,1,0}}%
      \expandafter\def\csname LT6\endcsname{\color[rgb]{0,0,0}}%
      \expandafter\def\csname LT7\endcsname{\color[rgb]{1,0.3,0}}%
      \expandafter\def\csname LT8\endcsname{\color[rgb]{0.5,0.5,0.5}}%
    \else
      \def\colorrgb#1{\color{black}}%
      \def\colorgray#1{\color[gray]{#1}}%
      \expandafter\def\csname LTw\endcsname{\color{white}}%
      \expandafter\def\csname LTb\endcsname{\color{black}}%
      \expandafter\def\csname LTa\endcsname{\color{black}}%
      \expandafter\def\csname LT0\endcsname{\color{black}}%
      \expandafter\def\csname LT1\endcsname{\color{black}}%
      \expandafter\def\csname LT2\endcsname{\color{black}}%
      \expandafter\def\csname LT3\endcsname{\color{black}}%
      \expandafter\def\csname LT4\endcsname{\color{black}}%
      \expandafter\def\csname LT5\endcsname{\color{black}}%
      \expandafter\def\csname LT6\endcsname{\color{black}}%
      \expandafter\def\csname LT7\endcsname{\color{black}}%
      \expandafter\def\csname LT8\endcsname{\color{black}}%
    \fi
  \fi
    \setlength{\unitlength}{0.0500bp}%
    \ifx\gptboxheight\undefined%
      \newlength{\gptboxheight}%
      \newlength{\gptboxwidth}%
      \newsavebox{\gptboxtext}%
    \fi%
    \setlength{\fboxrule}{0.5pt}%
    \setlength{\fboxsep}{1pt}%
\begin{picture}(3528.00,1440.00)%
    \gplgaddtomacro\gplbacktext{%
      \csname LTb\endcsname%
      \put(726,440){\makebox(0,0)[r]{\strut{}$-1$}}%
      \csname LTb\endcsname%
      \put(726,635){\makebox(0,0)[r]{\strut{}$-0.5$}}%
      \csname LTb\endcsname%
      \put(726,830){\makebox(0,0)[r]{\strut{}$0$}}%
      \csname LTb\endcsname%
      \put(726,1024){\makebox(0,0)[r]{\strut{}$0.5$}}%
      \csname LTb\endcsname%
      \put(726,1219){\makebox(0,0)[r]{\strut{}$1$}}%
      \put(2374,220){\makebox(0,0){\strut{}2013}}%
      \put(3131,220){\makebox(0,0){\strut{}2014}}%
      \put(858,220){\makebox(0,0){\strut{}2008}}%
      \put(1617,220){\makebox(0,0){\strut{}2009}}%
    }%
    \gplgaddtomacro\gplfronttext{%
      \csname LTb\endcsname%
      \put(220,829){\rotatebox{-270}{\makebox(0,0){\strut{}$O - C$, mas}}}%
      \put(1994,1439){\makebox(0,0){\strut{}Mars, VLBA, $\alpha$}}%
    }%
    \gplbacktext
    \put(0,0){\includegraphics{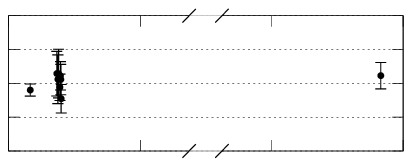}}%
    \gplfronttext
  \end{picture}%
\endgroup

%% file: vlba-mars-delta.tex
\begingroup
  \makeatletter
  \providecommand\color[2][]{%
    \GenericError{(gnuplot) \space\space\space\@spaces}{%
      Package color not loaded in conjunction with
      terminal option `colourtext'%
    }{See the gnuplot documentation for explanation.%
    }{Either use 'blacktext' in gnuplot or load the package
      color.sty in LaTeX.}%
    \renewcommand\color[2][]{}%
  }%
  \providecommand\includegraphics[2][]{%
    \GenericError{(gnuplot) \space\space\space\@spaces}{%
      Package graphicx or graphics not loaded%
    }{See the gnuplot documentation for explanation.%
    }{The gnuplot epslatex terminal needs graphicx.sty or graphics.sty.}%
    \renewcommand\includegraphics[2][]{}%
  }%
  \providecommand\rotatebox[2]{#2}%
  \@ifundefined{ifGPcolor}{%
    \newif\ifGPcolor
    \GPcolorfalse
  }{}%
  \@ifundefined{ifGPblacktext}{%
    \newif\ifGPblacktext
    \GPblacktexttrue
  }{}%
  \let\gplgaddtomacro\g@addto@macro
  \gdef\gplbacktext{}%
  \gdef\gplfronttext{}%
  \makeatother
  \ifGPblacktext
    \def\colorrgb#1{}%
    \def\colorgray#1{}%
  \else
    \ifGPcolor
      \def\colorrgb#1{\color[rgb]{#1}}%
      \def\colorgray#1{\color[gray]{#1}}%
      \expandafter\def\csname LTw\endcsname{\color{white}}%
      \expandafter\def\csname LTb\endcsname{\color{black}}%
      \expandafter\def\csname LTa\endcsname{\color{black}}%
      \expandafter\def\csname LT0\endcsname{\color[rgb]{1,0,0}}%
      \expandafter\def\csname LT1\endcsname{\color[rgb]{0,1,0}}%
      \expandafter\def\csname LT2\endcsname{\color[rgb]{0,0,1}}%
      \expandafter\def\csname LT3\endcsname{\color[rgb]{1,0,1}}%
      \expandafter\def\csname LT4\endcsname{\color[rgb]{0,1,1}}%
      \expandafter\def\csname LT5\endcsname{\color[rgb]{1,1,0}}%
      \expandafter\def\csname LT6\endcsname{\color[rgb]{0,0,0}}%
      \expandafter\def\csname LT7\endcsname{\color[rgb]{1,0.3,0}}%
      \expandafter\def\csname LT8\endcsname{\color[rgb]{0.5,0.5,0.5}}%
    \else
      \def\colorrgb#1{\color{black}}%
      \def\colorgray#1{\color[gray]{#1}}%
      \expandafter\def\csname LTw\endcsname{\color{white}}%
      \expandafter\def\csname LTb\endcsname{\color{black}}%
      \expandafter\def\csname LTa\endcsname{\color{black}}%
      \expandafter\def\csname LT0\endcsname{\color{black}}%
      \expandafter\def\csname LT1\endcsname{\color{black}}%
      \expandafter\def\csname LT2\endcsname{\color{black}}%
      \expandafter\def\csname LT3\endcsname{\color{black}}%
      \expandafter\def\csname LT4\endcsname{\color{black}}%
      \expandafter\def\csname LT5\endcsname{\color{black}}%
      \expandafter\def\csname LT6\endcsname{\color{black}}%
      \expandafter\def\csname LT7\endcsname{\color{black}}%
      \expandafter\def\csname LT8\endcsname{\color{black}}%
    \fi
  \fi
    \setlength{\unitlength}{0.0500bp}%
    \ifx\gptboxheight\undefined%
      \newlength{\gptboxheight}%
      \newlength{\gptboxwidth}%
      \newsavebox{\gptboxtext}%
    \fi%
    \setlength{\fboxrule}{0.5pt}%
    \setlength{\fboxsep}{1pt}%
\begin{picture}(3528.00,1440.00)%
    \gplgaddtomacro\gplbacktext{%
      \csname LTb\endcsname%
      \put(726,440){\makebox(0,0)[r]{\strut{}$-1$}}%
      \csname LTb\endcsname%
      \put(726,635){\makebox(0,0)[r]{\strut{}$-0.5$}}%
      \csname LTb\endcsname%
      \put(726,830){\makebox(0,0)[r]{\strut{}$0$}}%
      \csname LTb\endcsname%
      \put(726,1024){\makebox(0,0)[r]{\strut{}$0.5$}}%
      \csname LTb\endcsname%
      \put(726,1219){\makebox(0,0)[r]{\strut{}$1$}}%
      \put(2374,220){\makebox(0,0){\strut{}2013}}%
      \put(3131,220){\makebox(0,0){\strut{}2014}}%
      \put(858,220){\makebox(0,0){\strut{}2008}}%
      \put(1617,220){\makebox(0,0){\strut{}2009}}%
    }%
    \gplgaddtomacro\gplfronttext{%
      \csname LTb\endcsname%
      \put(220,829){\rotatebox{-270}{\makebox(0,0){\strut{}$O - C$, mas}}}%
      \put(1994,1439){\makebox(0,0){\strut{}Mars, VLBA, $\delta$}}%
    }%
    \gplbacktext
    \put(0,0){\includegraphics{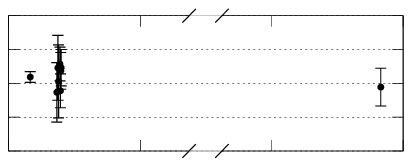}}%
    \gplfronttext
  \end{picture}%
\endgroup

%% file: ut0.tex
\begingroup
  \makeatletter
  \providecommand\color[2][]{%
    \GenericError{(gnuplot) \space\space\space\@spaces}{%
      Package color not loaded in conjunction with
      terminal option `colourtext'%
    }{See the gnuplot documentation for explanation.%
    }{Either use 'blacktext' in gnuplot or load the package
      color.sty in LaTeX.}%
    \renewcommand\color[2][]{}%
  }%
  \providecommand\includegraphics[2][]{%
    \GenericError{(gnuplot) \space\space\space\@spaces}{%
      Package graphicx or graphics not loaded%
    }{See the gnuplot documentation for explanation.%
    }{The gnuplot epslatex terminal needs graphicx.sty or graphics.sty.}%
    \renewcommand\includegraphics[2][]{}%
  }%
  \providecommand\rotatebox[2]{#2}%
  \@ifundefined{ifGPcolor}{%
    \newif\ifGPcolor
    \GPcolorfalse
  }{}%
  \@ifundefined{ifGPblacktext}{%
    \newif\ifGPblacktext
    \GPblacktexttrue
  }{}%
  \let\gplgaddtomacro\g@addto@macro
  \gdef\gplbacktext{}%
  \gdef\gplfronttext{}%
  \makeatother
  \ifGPblacktext
    \def\colorrgb#1{}%
    \def\colorgray#1{}%
  \else
    \ifGPcolor
      \def\colorrgb#1{\color[rgb]{#1}}%
      \def\colorgray#1{\color[gray]{#1}}%
      \expandafter\def\csname LTw\endcsname{\color{white}}%
      \expandafter\def\csname LTb\endcsname{\color{black}}%
      \expandafter\def\csname LTa\endcsname{\color{black}}%
      \expandafter\def\csname LT0\endcsname{\color[rgb]{1,0,0}}%
      \expandafter\def\csname LT1\endcsname{\color[rgb]{0,1,0}}%
      \expandafter\def\csname LT2\endcsname{\color[rgb]{0,0,1}}%
      \expandafter\def\csname LT3\endcsname{\color[rgb]{1,0,1}}%
      \expandafter\def\csname LT4\endcsname{\color[rgb]{0,1,1}}%
      \expandafter\def\csname LT5\endcsname{\color[rgb]{1,1,0}}%
      \expandafter\def\csname LT6\endcsname{\color[rgb]{0,0,0}}%
      \expandafter\def\csname LT7\endcsname{\color[rgb]{1,0.3,0}}%
      \expandafter\def\csname LT8\endcsname{\color[rgb]{0.5,0.5,0.5}}%
    \else
      \def\colorrgb#1{\color{black}}%
      \def\colorgray#1{\color[gray]{#1}}%
      \expandafter\def\csname LTw\endcsname{\color{white}}%
      \expandafter\def\csname LTb\endcsname{\color{black}}%
      \expandafter\def\csname LTa\endcsname{\color{black}}%
      \expandafter\def\csname LT0\endcsname{\color{black}}%
      \expandafter\def\csname LT1\endcsname{\color{black}}%
      \expandafter\def\csname LT2\endcsname{\color{black}}%
      \expandafter\def\csname LT3\endcsname{\color{black}}%
      \expandafter\def\csname LT4\endcsname{\color{black}}%
      \expandafter\def\csname LT5\endcsname{\color{black}}%
      \expandafter\def\csname LT6\endcsname{\color{black}}%
      \expandafter\def\csname LT7\endcsname{\color{black}}%
      \expandafter\def\csname LT8\endcsname{\color{black}}%
    \fi
  \fi
    \setlength{\unitlength}{0.0500bp}%
    \ifx\gptboxheight\undefined%
      \newlength{\gptboxheight}%
      \newlength{\gptboxwidth}%
      \newsavebox{\gptboxtext}%
    \fi%
    \setlength{\fboxrule}{0.5pt}%
    \setlength{\fboxsep}{1pt}%
\begin{picture}(5040.00,1872.00)%
    \gplgaddtomacro\gplbacktext{%
      \csname LTb\endcsname%
      \put(616,440){\makebox(0,0)[r]{\strut{}$0$}}%
      \csname LTb\endcsname%
      \put(616,743){\makebox(0,0)[r]{\strut{}$0.1$}}%
      \csname LTb\endcsname%
      \put(616,1046){\makebox(0,0)[r]{\strut{}$0.2$}}%
      \csname LTb\endcsname%
      \put(616,1348){\makebox(0,0)[r]{\strut{}$0.3$}}%
      \csname LTb\endcsname%
      \put(616,1651){\makebox(0,0)[r]{\strut{}$0.4$}}%
      \put(752,220){\makebox(0,0){\strut{}2014}}%
      \put(1449,220){\makebox(0,0){\strut{}2015}}%
      \put(2146,220){\makebox(0,0){\strut{}2016}}%
      \put(2844,220){\makebox(0,0){\strut{}2017}}%
      \put(3541,220){\makebox(0,0){\strut{}2018}}%
      \put(4238,220){\makebox(0,0){\strut{}2019}}%
    }%
    \gplgaddtomacro\gplfronttext{%
      \csname LTb\endcsname%
      \put(176,1045){\rotatebox{-270}{\makebox(0,0){\strut{}$\sigma(\Delta\mathrm{UT0})$, ms}}}%
      \csname LTb\endcsname%
      \put(1870,1754){\makebox(0,0)[r]{\strut{}Grasse (IR)      }}%
      \csname LTb\endcsname%
      \put(3187,1754){\makebox(0,0)[r]{\strut{}Grasse (MeO)     }}%
      \csname LTb\endcsname%
      \put(4504,1754){\makebox(0,0)[r]{\strut{}Apache Point     }}%
    }%
    \gplbacktext
    \put(0,0){\includegraphics{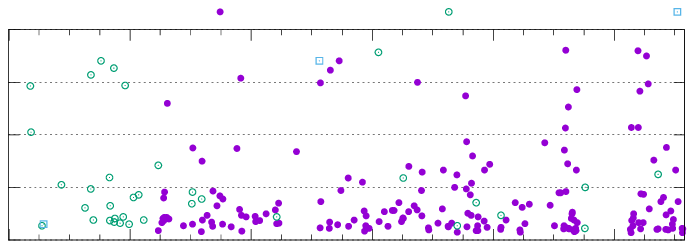}}%
    \gplfronttext
  \end{picture}%
\endgroup

%% file: vol.tex
\begingroup
  \makeatletter
  \providecommand\color[2][]{%
    \GenericError{(gnuplot) \space\space\space\@spaces}{%
      Package color not loaded in conjunction with
      terminal option `colourtext'%
    }{See the gnuplot documentation for explanation.%
    }{Either use 'blacktext' in gnuplot or load the package
      color.sty in LaTeX.}%
    \renewcommand\color[2][]{}%
  }%
  \providecommand\includegraphics[2][]{%
    \GenericError{(gnuplot) \space\space\space\@spaces}{%
      Package graphicx or graphics not loaded%
    }{See the gnuplot documentation for explanation.%
    }{The gnuplot epslatex terminal needs graphicx.sty or graphics.sty.}%
    \renewcommand\includegraphics[2][]{}%
  }%
  \providecommand\rotatebox[2]{#2}%
  \@ifundefined{ifGPcolor}{%
    \newif\ifGPcolor
    \GPcolorfalse
  }{}%
  \@ifundefined{ifGPblacktext}{%
    \newif\ifGPblacktext
    \GPblacktexttrue
  }{}%
  \let\gplgaddtomacro\g@addto@macro
  \gdef\gplbacktext{}%
  \gdef\gplfronttext{}%
  \makeatother
  \ifGPblacktext
    \def\colorrgb#1{}%
    \def\colorgray#1{}%
  \else
    \ifGPcolor
      \def\colorrgb#1{\color[rgb]{#1}}%
      \def\colorgray#1{\color[gray]{#1}}%
      \expandafter\def\csname LTw\endcsname{\color{white}}%
      \expandafter\def\csname LTb\endcsname{\color{black}}%
      \expandafter\def\csname LTa\endcsname{\color{black}}%
      \expandafter\def\csname LT0\endcsname{\color[rgb]{1,0,0}}%
      \expandafter\def\csname LT1\endcsname{\color[rgb]{0,1,0}}%
      \expandafter\def\csname LT2\endcsname{\color[rgb]{0,0,1}}%
      \expandafter\def\csname LT3\endcsname{\color[rgb]{1,0,1}}%
      \expandafter\def\csname LT4\endcsname{\color[rgb]{0,1,1}}%
      \expandafter\def\csname LT5\endcsname{\color[rgb]{1,1,0}}%
      \expandafter\def\csname LT6\endcsname{\color[rgb]{0,0,0}}%
      \expandafter\def\csname LT7\endcsname{\color[rgb]{1,0.3,0}}%
      \expandafter\def\csname LT8\endcsname{\color[rgb]{0.5,0.5,0.5}}%
    \else
      \def\colorrgb#1{\color{black}}%
      \def\colorgray#1{\color[gray]{#1}}%
      \expandafter\def\csname LTw\endcsname{\color{white}}%
      \expandafter\def\csname LTb\endcsname{\color{black}}%
      \expandafter\def\csname LTa\endcsname{\color{black}}%
      \expandafter\def\csname LT0\endcsname{\color{black}}%
      \expandafter\def\csname LT1\endcsname{\color{black}}%
      \expandafter\def\csname LT2\endcsname{\color{black}}%
      \expandafter\def\csname LT3\endcsname{\color{black}}%
      \expandafter\def\csname LT4\endcsname{\color{black}}%
      \expandafter\def\csname LT5\endcsname{\color{black}}%
      \expandafter\def\csname LT6\endcsname{\color{black}}%
      \expandafter\def\csname LT7\endcsname{\color{black}}%
      \expandafter\def\csname LT8\endcsname{\color{black}}%
    \fi
  \fi
    \setlength{\unitlength}{0.0500bp}%
    \ifx\gptboxheight\undefined%
      \newlength{\gptboxheight}%
      \newlength{\gptboxwidth}%
      \newsavebox{\gptboxtext}%
    \fi%
    \setlength{\fboxrule}{0.5pt}%
    \setlength{\fboxsep}{1pt}%
\begin{picture}(5040.00,1872.00)%
    \gplgaddtomacro\gplbacktext{%
      \csname LTb\endcsname%
      \put(573,440){\makebox(0,0)[r]{\strut{}$0$}}%
      \csname LTb\endcsname%
      \put(573,642){\makebox(0,0)[r]{\strut{}$1$}}%
      \csname LTb\endcsname%
      \put(573,844){\makebox(0,0)[r]{\strut{}$2$}}%
      \csname LTb\endcsname%
      \put(573,1046){\makebox(0,0)[r]{\strut{}$3$}}%
      \csname LTb\endcsname%
      \put(573,1247){\makebox(0,0)[r]{\strut{}$4$}}%
      \csname LTb\endcsname%
      \put(573,1449){\makebox(0,0)[r]{\strut{}$5$}}%
      \csname LTb\endcsname%
      \put(573,1651){\makebox(0,0)[r]{\strut{}$6$}}%
      \put(709,220){\makebox(0,0){\strut{}2014}}%
      \put(1413,220){\makebox(0,0){\strut{}2015}}%
      \put(2118,220){\makebox(0,0){\strut{}2016}}%
      \put(2825,220){\makebox(0,0){\strut{}2017}}%
      \put(3529,220){\makebox(0,0){\strut{}2018}}%
      \put(4234,220){\makebox(0,0){\strut{}2019}}%
    }%
    \gplgaddtomacro\gplfronttext{%
      \csname LTb\endcsname%
      \put(397,1045){\rotatebox{-270}{\makebox(0,0){\strut{}$\sigma(\mathrm{VOL})$, mas}}}%
    }%
    \gplbacktext
    \put(0,0){\includegraphics{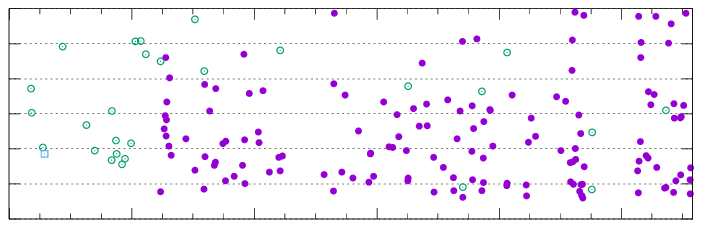}}%
    \gplfronttext
  \end{picture}%
\endgroup

%% file: ut0-corr-c04.tex
\begingroup
  \makeatletter
  \providecommand\color[2][]{%
    \GenericError{(gnuplot) \space\space\space\@spaces}{%
      Package color not loaded in conjunction with
      terminal option `colourtext'%
    }{See the gnuplot documentation for explanation.%
    }{Either use 'blacktext' in gnuplot or load the package
      color.sty in LaTeX.}%
    \renewcommand\color[2][]{}%
  }%
  \providecommand\includegraphics[2][]{%
    \GenericError{(gnuplot) \space\space\space\@spaces}{%
      Package graphicx or graphics not loaded%
    }{See the gnuplot documentation for explanation.%
    }{The gnuplot epslatex terminal needs graphicx.sty or graphics.sty.}%
    \renewcommand\includegraphics[2][]{}%
  }%
  \providecommand\rotatebox[2]{#2}%
  \@ifundefined{ifGPcolor}{%
    \newif\ifGPcolor
    \GPcolorfalse
  }{}%
  \@ifundefined{ifGPblacktext}{%
    \newif\ifGPblacktext
    \GPblacktexttrue
  }{}%
  \let\gplgaddtomacro\g@addto@macro
  \gdef\gplbacktext{}%
  \gdef\gplfronttext{}%
  \makeatother
  \ifGPblacktext
    \def\colorrgb#1{}%
    \def\colorgray#1{}%
  \else
    \ifGPcolor
      \def\colorrgb#1{\color[rgb]{#1}}%
      \def\colorgray#1{\color[gray]{#1}}%
      \expandafter\def\csname LTw\endcsname{\color{white}}%
      \expandafter\def\csname LTb\endcsname{\color{black}}%
      \expandafter\def\csname LTa\endcsname{\color{black}}%
      \expandafter\def\csname LT0\endcsname{\color[rgb]{1,0,0}}%
      \expandafter\def\csname LT1\endcsname{\color[rgb]{0,1,0}}%
      \expandafter\def\csname LT2\endcsname{\color[rgb]{0,0,1}}%
      \expandafter\def\csname LT3\endcsname{\color[rgb]{1,0,1}}%
      \expandafter\def\csname LT4\endcsname{\color[rgb]{0,1,1}}%
      \expandafter\def\csname LT5\endcsname{\color[rgb]{1,1,0}}%
      \expandafter\def\csname LT6\endcsname{\color[rgb]{0,0,0}}%
      \expandafter\def\csname LT7\endcsname{\color[rgb]{1,0.3,0}}%
      \expandafter\def\csname LT8\endcsname{\color[rgb]{0.5,0.5,0.5}}%
    \else
      \def\colorrgb#1{\color{black}}%
      \def\colorgray#1{\color[gray]{#1}}%
      \expandafter\def\csname LTw\endcsname{\color{white}}%
      \expandafter\def\csname LTb\endcsname{\color{black}}%
      \expandafter\def\csname LTa\endcsname{\color{black}}%
      \expandafter\def\csname LT0\endcsname{\color{black}}%
      \expandafter\def\csname LT1\endcsname{\color{black}}%
      \expandafter\def\csname LT2\endcsname{\color{black}}%
      \expandafter\def\csname LT3\endcsname{\color{black}}%
      \expandafter\def\csname LT4\endcsname{\color{black}}%
      \expandafter\def\csname LT5\endcsname{\color{black}}%
      \expandafter\def\csname LT6\endcsname{\color{black}}%
      \expandafter\def\csname LT7\endcsname{\color{black}}%
      \expandafter\def\csname LT8\endcsname{\color{black}}%
    \fi
  \fi
    \setlength{\unitlength}{0.0500bp}%
    \ifx\gptboxheight\undefined%
      \newlength{\gptboxheight}%
      \newlength{\gptboxwidth}%
      \newsavebox{\gptboxtext}%
    \fi%
    \setlength{\fboxrule}{0.5pt}%
    \setlength{\fboxsep}{1pt}%
\begin{picture}(3310.00,1872.00)%
    \gplgaddtomacro\gplbacktext{%
      \csname LTb\endcsname%
      \put(861,440){\makebox(0,0)[r]{\strut{}$-0.5$}}%
      \csname LTb\endcsname%
      \put(861,613){\makebox(0,0)[r]{\strut{}$-0.4$}}%
      \csname LTb\endcsname%
      \put(861,786){\makebox(0,0)[r]{\strut{}$-0.3$}}%
      \csname LTb\endcsname%
      \put(861,959){\makebox(0,0)[r]{\strut{}$-0.2$}}%
      \csname LTb\endcsname%
      \put(861,1132){\makebox(0,0)[r]{\strut{}$-0.1$}}%
      \csname LTb\endcsname%
      \put(861,1305){\makebox(0,0)[r]{\strut{}$0$}}%
      \csname LTb\endcsname%
      \put(861,1478){\makebox(0,0)[r]{\strut{}$0.1$}}%
      \csname LTb\endcsname%
      \put(861,1651){\makebox(0,0)[r]{\strut{}$0.2$}}%
      \put(997,220){\makebox(0,0){\strut{}2017}}%
      \put(1739,220){\makebox(0,0){\strut{}2018}}%
      \put(2482,220){\makebox(0,0){\strut{}2019}}%
    }%
    \gplgaddtomacro\gplfronttext{%
      \csname LTb\endcsname%
      \put(289,1045){\rotatebox{-270}{\makebox(0,0){\strut{}$\Delta$UT0, ms}}}%
    }%
    \gplbacktext
    \put(0,0){\includegraphics{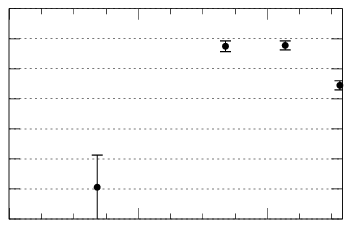}}%
    \gplfronttext
  \end{picture}%
\endgroup

%% file: ut0-corr-finals.tex
\begingroup
  \makeatletter
  \providecommand\color[2][]{%
    \GenericError{(gnuplot) \space\space\space\@spaces}{%
      Package color not loaded in conjunction with
      terminal option `colourtext'%
    }{See the gnuplot documentation for explanation.%
    }{Either use 'blacktext' in gnuplot or load the package
      color.sty in LaTeX.}%
    \renewcommand\color[2][]{}%
  }%
  \providecommand\includegraphics[2][]{%
    \GenericError{(gnuplot) \space\space\space\@spaces}{%
      Package graphicx or graphics not loaded%
    }{See the gnuplot documentation for explanation.%
    }{The gnuplot epslatex terminal needs graphicx.sty or graphics.sty.}%
    \renewcommand\includegraphics[2][]{}%
  }%
  \providecommand\rotatebox[2]{#2}%
  \@ifundefined{ifGPcolor}{%
    \newif\ifGPcolor
    \GPcolorfalse
  }{}%
  \@ifundefined{ifGPblacktext}{%
    \newif\ifGPblacktext
    \GPblacktexttrue
  }{}%
  \let\gplgaddtomacro\g@addto@macro
  \gdef\gplbacktext{}%
  \gdef\gplfronttext{}%
  \makeatother
  \ifGPblacktext
    \def\colorrgb#1{}%
    \def\colorgray#1{}%
  \else
    \ifGPcolor
      \def\colorrgb#1{\color[rgb]{#1}}%
      \def\colorgray#1{\color[gray]{#1}}%
      \expandafter\def\csname LTw\endcsname{\color{white}}%
      \expandafter\def\csname LTb\endcsname{\color{black}}%
      \expandafter\def\csname LTa\endcsname{\color{black}}%
      \expandafter\def\csname LT0\endcsname{\color[rgb]{1,0,0}}%
      \expandafter\def\csname LT1\endcsname{\color[rgb]{0,1,0}}%
      \expandafter\def\csname LT2\endcsname{\color[rgb]{0,0,1}}%
      \expandafter\def\csname LT3\endcsname{\color[rgb]{1,0,1}}%
      \expandafter\def\csname LT4\endcsname{\color[rgb]{0,1,1}}%
      \expandafter\def\csname LT5\endcsname{\color[rgb]{1,1,0}}%
      \expandafter\def\csname LT6\endcsname{\color[rgb]{0,0,0}}%
      \expandafter\def\csname LT7\endcsname{\color[rgb]{1,0.3,0}}%
      \expandafter\def\csname LT8\endcsname{\color[rgb]{0.5,0.5,0.5}}%
    \else
      \def\colorrgb#1{\color{black}}%
      \def\colorgray#1{\color[gray]{#1}}%
      \expandafter\def\csname LTw\endcsname{\color{white}}%
      \expandafter\def\csname LTb\endcsname{\color{black}}%
      \expandafter\def\csname LTa\endcsname{\color{black}}%
      \expandafter\def\csname LT0\endcsname{\color{black}}%
      \expandafter\def\csname LT1\endcsname{\color{black}}%
      \expandafter\def\csname LT2\endcsname{\color{black}}%
      \expandafter\def\csname LT3\endcsname{\color{black}}%
      \expandafter\def\csname LT4\endcsname{\color{black}}%
      \expandafter\def\csname LT5\endcsname{\color{black}}%
      \expandafter\def\csname LT6\endcsname{\color{black}}%
      \expandafter\def\csname LT7\endcsname{\color{black}}%
      \expandafter\def\csname LT8\endcsname{\color{black}}%
    \fi
  \fi
    \setlength{\unitlength}{0.0500bp}%
    \ifx\gptboxheight\undefined%
      \newlength{\gptboxheight}%
      \newlength{\gptboxwidth}%
      \newsavebox{\gptboxtext}%
    \fi%
    \setlength{\fboxrule}{0.5pt}%
    \setlength{\fboxsep}{1pt}%
\begin{picture}(3310.00,1872.00)%
    \gplgaddtomacro\gplbacktext{%
      \csname LTb\endcsname%
      \put(861,440){\makebox(0,0)[r]{\strut{}$-0.5$}}%
      \csname LTb\endcsname%
      \put(861,613){\makebox(0,0)[r]{\strut{}$-0.4$}}%
      \csname LTb\endcsname%
      \put(861,786){\makebox(0,0)[r]{\strut{}$-0.3$}}%
      \csname LTb\endcsname%
      \put(861,959){\makebox(0,0)[r]{\strut{}$-0.2$}}%
      \csname LTb\endcsname%
      \put(861,1132){\makebox(0,0)[r]{\strut{}$-0.1$}}%
      \csname LTb\endcsname%
      \put(861,1305){\makebox(0,0)[r]{\strut{}$0$}}%
      \csname LTb\endcsname%
      \put(861,1478){\makebox(0,0)[r]{\strut{}$0.1$}}%
      \csname LTb\endcsname%
      \put(861,1651){\makebox(0,0)[r]{\strut{}$0.2$}}%
      \put(997,220){\makebox(0,0){\strut{}2017}}%
      \put(1739,220){\makebox(0,0){\strut{}2018}}%
      \put(2482,220){\makebox(0,0){\strut{}2019}}%
    }%
    \gplgaddtomacro\gplfronttext{%
      \csname LTb\endcsname%
      \put(289,1045){\rotatebox{-270}{\makebox(0,0){\strut{}$\Delta$UT0, ms}}}%
    }%
    \gplbacktext
    \put(0,0){\includegraphics{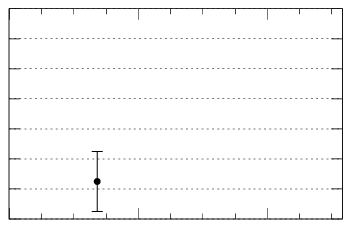}}%
    \gplfronttext
  \end{picture}%
\endgroup

%% file: lop-r.tex
\begingroup
  \makeatletter
  \providecommand\color[2][]{%
    \GenericError{(gnuplot) \space\space\space\@spaces}{%
      Package color not loaded in conjunction with
      terminal option `colourtext'%
    }{See the gnuplot documentation for explanation.%
    }{Either use 'blacktext' in gnuplot or load the package
      color.sty in LaTeX.}%
    \renewcommand\color[2][]{}%
  }%
  \providecommand\includegraphics[2][]{%
    \GenericError{(gnuplot) \space\space\space\@spaces}{%
      Package graphicx or graphics not loaded%
    }{See the gnuplot documentation for explanation.%
    }{The gnuplot epslatex terminal needs graphicx.sty or graphics.sty.}%
    \renewcommand\includegraphics[2][]{}%
  }%
  \providecommand\rotatebox[2]{#2}%
  \@ifundefined{ifGPcolor}{%
    \newif\ifGPcolor
    \GPcolorfalse
  }{}%
  \@ifundefined{ifGPblacktext}{%
    \newif\ifGPblacktext
    \GPblacktexttrue
  }{}%
  \let\gplgaddtomacro\g@addto@macro
  \gdef\gplbacktext{}%
  \gdef\gplfronttext{}%
  \makeatother
  \ifGPblacktext
    \def\colorrgb#1{}%
    \def\colorgray#1{}%
  \else
    \ifGPcolor
      \def\colorrgb#1{\color[rgb]{#1}}%
      \def\colorgray#1{\color[gray]{#1}}%
      \expandafter\def\csname LTw\endcsname{\color{white}}%
      \expandafter\def\csname LTb\endcsname{\color{black}}%
      \expandafter\def\csname LTa\endcsname{\color{black}}%
      \expandafter\def\csname LT0\endcsname{\color[rgb]{1,0,0}}%
      \expandafter\def\csname LT1\endcsname{\color[rgb]{0,1,0}}%
      \expandafter\def\csname LT2\endcsname{\color[rgb]{0,0,1}}%
      \expandafter\def\csname LT3\endcsname{\color[rgb]{1,0,1}}%
      \expandafter\def\csname LT4\endcsname{\color[rgb]{0,1,1}}%
      \expandafter\def\csname LT5\endcsname{\color[rgb]{1,1,0}}%
      \expandafter\def\csname LT6\endcsname{\color[rgb]{0,0,0}}%
      \expandafter\def\csname LT7\endcsname{\color[rgb]{1,0.3,0}}%
      \expandafter\def\csname LT8\endcsname{\color[rgb]{0.5,0.5,0.5}}%
    \else
      \def\colorrgb#1{\color{black}}%
      \def\colorgray#1{\color[gray]{#1}}%
      \expandafter\def\csname LTw\endcsname{\color{white}}%
      \expandafter\def\csname LTb\endcsname{\color{black}}%
      \expandafter\def\csname LTa\endcsname{\color{black}}%
      \expandafter\def\csname LT0\endcsname{\color{black}}%
      \expandafter\def\csname LT1\endcsname{\color{black}}%
      \expandafter\def\csname LT2\endcsname{\color{black}}%
      \expandafter\def\csname LT3\endcsname{\color{black}}%
      \expandafter\def\csname LT4\endcsname{\color{black}}%
      \expandafter\def\csname LT5\endcsname{\color{black}}%
      \expandafter\def\csname LT6\endcsname{\color{black}}%
      \expandafter\def\csname LT7\endcsname{\color{black}}%
      \expandafter\def\csname LT8\endcsname{\color{black}}%
    \fi
  \fi
    \setlength{\unitlength}{0.0500bp}%
    \ifx\gptboxheight\undefined%
      \newlength{\gptboxheight}%
      \newlength{\gptboxwidth}%
      \newsavebox{\gptboxtext}%
    \fi%
    \setlength{\fboxrule}{0.5pt}%
    \setlength{\fboxsep}{1pt}%
\begin{picture}(5040.00,1872.00)%
    \gplgaddtomacro\gplbacktext{%
      \csname LTb\endcsname%
      \put(573,440){\makebox(0,0)[r]{\strut{}$0$}}%
      \csname LTb\endcsname%
      \put(573,642){\makebox(0,0)[r]{\strut{}$1$}}%
      \csname LTb\endcsname%
      \put(573,844){\makebox(0,0)[r]{\strut{}$2$}}%
      \csname LTb\endcsname%
      \put(573,1046){\makebox(0,0)[r]{\strut{}$3$}}%
      \csname LTb\endcsname%
      \put(573,1247){\makebox(0,0)[r]{\strut{}$4$}}%
      \csname LTb\endcsname%
      \put(573,1449){\makebox(0,0)[r]{\strut{}$5$}}%
      \csname LTb\endcsname%
      \put(573,1651){\makebox(0,0)[r]{\strut{}$6$}}%
      \put(709,220){\makebox(0,0){\strut{}2014}}%
      \put(1413,220){\makebox(0,0){\strut{}2015}}%
      \put(2118,220){\makebox(0,0){\strut{}2016}}%
      \put(2825,220){\makebox(0,0){\strut{}2017}}%
      \put(3529,220){\makebox(0,0){\strut{}2018}}%
      \put(4234,220){\makebox(0,0){\strut{}2019}}%
    }%
    \gplgaddtomacro\gplfronttext{%
      \csname LTb\endcsname%
      \put(397,1045){\rotatebox{-270}{\makebox(0,0){\strut{}$\sigma(r')$, mas}}}%
      \csname LTb\endcsname%
      \put(1827,1754){\makebox(0,0)[r]{\strut{}Grasse (IR)     }}%
      \csname LTb\endcsname%
      \put(3144,1754){\makebox(0,0)[r]{\strut{}Grasse (MeO)    }}%
      \csname LTb\endcsname%
      \put(4461,1754){\makebox(0,0)[r]{\strut{}Apache Point     }}%
    }%
    \gplbacktext
    \put(0,0){\includegraphics{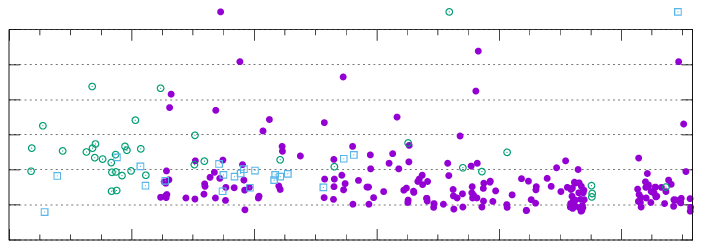}}%
    \gplfronttext
  \end{picture}%
\endgroup

%% file: lop-q.tex
\begingroup
  \makeatletter
  \providecommand\color[2][]{%
    \GenericError{(gnuplot) \space\space\space\@spaces}{%
      Package color not loaded in conjunction with
      terminal option `colourtext'%
    }{See the gnuplot documentation for explanation.%
    }{Either use 'blacktext' in gnuplot or load the package
      color.sty in LaTeX.}%
    \renewcommand\color[2][]{}%
  }%
  \providecommand\includegraphics[2][]{%
    \GenericError{(gnuplot) \space\space\space\@spaces}{%
      Package graphicx or graphics not loaded%
    }{See the gnuplot documentation for explanation.%
    }{The gnuplot epslatex terminal needs graphicx.sty or graphics.sty.}%
    \renewcommand\includegraphics[2][]{}%
  }%
  \providecommand\rotatebox[2]{#2}%
  \@ifundefined{ifGPcolor}{%
    \newif\ifGPcolor
    \GPcolorfalse
  }{}%
  \@ifundefined{ifGPblacktext}{%
    \newif\ifGPblacktext
    \GPblacktexttrue
  }{}%
  \let\gplgaddtomacro\g@addto@macro
  \gdef\gplbacktext{}%
  \gdef\gplfronttext{}%
  \makeatother
  \ifGPblacktext
    \def\colorrgb#1{}%
    \def\colorgray#1{}%
  \else
    \ifGPcolor
      \def\colorrgb#1{\color[rgb]{#1}}%
      \def\colorgray#1{\color[gray]{#1}}%
      \expandafter\def\csname LTw\endcsname{\color{white}}%
      \expandafter\def\csname LTb\endcsname{\color{black}}%
      \expandafter\def\csname LTa\endcsname{\color{black}}%
      \expandafter\def\csname LT0\endcsname{\color[rgb]{1,0,0}}%
      \expandafter\def\csname LT1\endcsname{\color[rgb]{0,1,0}}%
      \expandafter\def\csname LT2\endcsname{\color[rgb]{0,0,1}}%
      \expandafter\def\csname LT3\endcsname{\color[rgb]{1,0,1}}%
      \expandafter\def\csname LT4\endcsname{\color[rgb]{0,1,1}}%
      \expandafter\def\csname LT5\endcsname{\color[rgb]{1,1,0}}%
      \expandafter\def\csname LT6\endcsname{\color[rgb]{0,0,0}}%
      \expandafter\def\csname LT7\endcsname{\color[rgb]{1,0.3,0}}%
      \expandafter\def\csname LT8\endcsname{\color[rgb]{0.5,0.5,0.5}}%
    \else
      \def\colorrgb#1{\color{black}}%
      \def\colorgray#1{\color[gray]{#1}}%
      \expandafter\def\csname LTw\endcsname{\color{white}}%
      \expandafter\def\csname LTb\endcsname{\color{black}}%
      \expandafter\def\csname LTa\endcsname{\color{black}}%
      \expandafter\def\csname LT0\endcsname{\color{black}}%
      \expandafter\def\csname LT1\endcsname{\color{black}}%
      \expandafter\def\csname LT2\endcsname{\color{black}}%
      \expandafter\def\csname LT3\endcsname{\color{black}}%
      \expandafter\def\csname LT4\endcsname{\color{black}}%
      \expandafter\def\csname LT5\endcsname{\color{black}}%
      \expandafter\def\csname LT6\endcsname{\color{black}}%
      \expandafter\def\csname LT7\endcsname{\color{black}}%
      \expandafter\def\csname LT8\endcsname{\color{black}}%
    \fi
  \fi
    \setlength{\unitlength}{0.0500bp}%
    \ifx\gptboxheight\undefined%
      \newlength{\gptboxheight}%
      \newlength{\gptboxwidth}%
      \newsavebox{\gptboxtext}%
    \fi%
    \setlength{\fboxrule}{0.5pt}%
    \setlength{\fboxsep}{1pt}%
\begin{picture}(5040.00,1872.00)%
    \gplgaddtomacro\gplbacktext{%
      \csname LTb\endcsname%
      \put(573,440){\makebox(0,0)[r]{\strut{}$0$}}%
      \csname LTb\endcsname%
      \put(573,642){\makebox(0,0)[r]{\strut{}$1$}}%
      \csname LTb\endcsname%
      \put(573,844){\makebox(0,0)[r]{\strut{}$2$}}%
      \csname LTb\endcsname%
      \put(573,1046){\makebox(0,0)[r]{\strut{}$3$}}%
      \csname LTb\endcsname%
      \put(573,1247){\makebox(0,0)[r]{\strut{}$4$}}%
      \csname LTb\endcsname%
      \put(573,1449){\makebox(0,0)[r]{\strut{}$5$}}%
      \csname LTb\endcsname%
      \put(573,1651){\makebox(0,0)[r]{\strut{}$6$}}%
      \put(709,220){\makebox(0,0){\strut{}2014}}%
      \put(1413,220){\makebox(0,0){\strut{}2015}}%
      \put(2118,220){\makebox(0,0){\strut{}2016}}%
      \put(2825,220){\makebox(0,0){\strut{}2017}}%
      \put(3529,220){\makebox(0,0){\strut{}2018}}%
      \put(4234,220){\makebox(0,0){\strut{}2019}}%
    }%
    \gplgaddtomacro\gplfronttext{%
      \csname LTb\endcsname%
      \put(397,1045){\rotatebox{-270}{\makebox(0,0){\strut{}$\sigma(q')$, mas}}}%
    }%
    \gplbacktext
    \put(0,0){\includegraphics{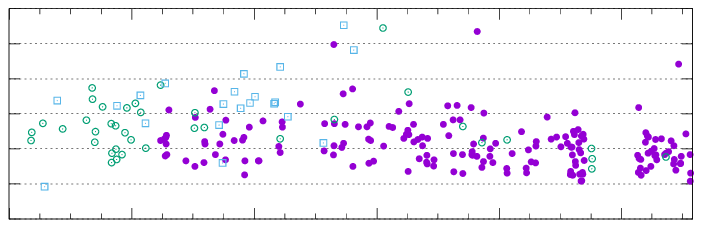}}%
    \gplfronttext
  \end{picture}%
\endgroup

%% file: ties.tex
\begingroup%
  \makeatletter%
  \providecommand\color[2][]{%
    \errmessage{(Inkscape) Color is used for the text in Inkscape, but the package 'color.sty' is not loaded}%
    \renewcommand\color[2][]{}%
  }%
  \providecommand\transparent[1]{%
    \errmessage{(Inkscape) Transparency is used (non-zero) for the text in Inkscape, but the package 'transparent.sty' is not loaded}%
    \renewcommand\transparent[1]{}%
  }%
  \providecommand\rotatebox[2]{#2}%
  \ifx\svgwidth\undefined%
    \setlength{\unitlength}{219.95943138bp}%
    \ifx\svgscale\undefined%
      \relax%
    \else%
      \setlength{\unitlength}{\unitlength * \real{\svgscale}}%
    \fi%
  \else%
    \setlength{\unitlength}{\svgwidth}%
  \fi%
  \global\let\svgwidth\undefined%
  \global\let\svgscale\undefined%
  \makeatother%
  \begin{picture}(1,0.44716361)%
    \put(0,0){\includegraphics[width=\unitlength]{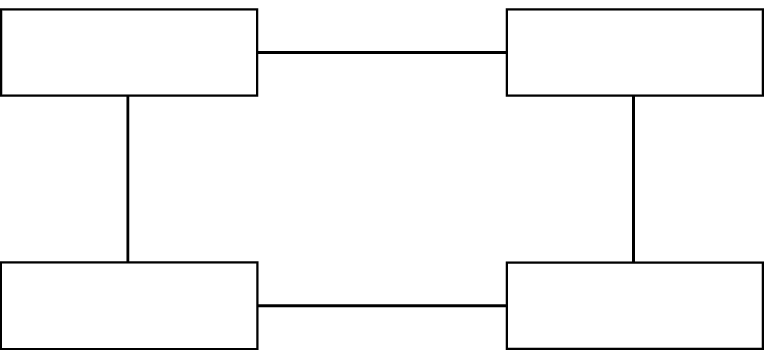}}%
    \put(0.10662198,0.378505){\color[rgb]{0,0,0}\makebox(0,0)[lb]{\smash{Ecliptic}}}%
    \put(1.38857264,2.88220421){\color[rgb]{0,0,0}\makebox(0,0)[lt]{\begin{minipage}{0.00827342\unitlength}\raggedright \end{minipage}}}%
    \put(0.02832829,0.07147783){\color[rgb]{0,0,0}\makebox(0,0)[lb]{\smash{Orbits of planets}}}%
    \put(0.02983499,0.02852021){\color[rgb]{0,0,0}\makebox(0,0)[lb]{\smash{other than Earth}}}%
    \put(0.35686455,0.07833749){\color[rgb]{0,0,0}\makebox(0,0)[lb]{\smash{Spacecraft VLBI}}}%
    \put(0.18704209,0.2326165){\color[rgb]{0,0,0}\makebox(0,0)[lb]{\smash{Spacecraft}}}%
    \put(0.76787855,0.38005002){\color[rgb]{0,0,0}\makebox(0,0)[lb]{\smash{Equator}}}%
    \put(0.78458693,0.04846757){\color[rgb]{0,0,0}\makebox(0,0)[lb]{\smash{ICRF}}}%
    \put(0.39802697,0.35352576){\color[rgb]{0,0,0}\makebox(0,0)[lb]{\smash{Lunar laser}}}%
    \put(0.39804875,0.31643358){\color[rgb]{0,0,0}\makebox(0,0)[lb]{\smash{ranging}}}%
    \put(0.18777692,0.19632622){\color[rgb]{0,0,0}\makebox(0,0)[lb]{\smash{ranging}}}%
    \put(0.72267907,0.2165772){\color[rgb]{0,0,0}\makebox(0,0)[lb]{\smash{VLBI}}}%
  \end{picture}%
\endgroup%

%% file: icrf-x-c04.tex
\begingroup
  \makeatletter
  \providecommand\color[2][]{%
    \GenericError{(gnuplot) \space\space\space\@spaces}{%
      Package color not loaded in conjunction with
      terminal option `colourtext'%
    }{See the gnuplot documentation for explanation.%
    }{Either use 'blacktext' in gnuplot or load the package
      color.sty in LaTeX.}%
    \renewcommand\color[2][]{}%
  }%
  \providecommand\includegraphics[2][]{%
    \GenericError{(gnuplot) \space\space\space\@spaces}{%
      Package graphicx or graphics not loaded%
    }{See the gnuplot documentation for explanation.%
    }{The gnuplot epslatex terminal needs graphicx.sty or graphics.sty.}%
    \renewcommand\includegraphics[2][]{}%
  }%
  \providecommand\rotatebox[2]{#2}%
  \@ifundefined{ifGPcolor}{%
    \newif\ifGPcolor
    \GPcolorfalse
  }{}%
  \@ifundefined{ifGPblacktext}{%
    \newif\ifGPblacktext
    \GPblacktexttrue
  }{}%
  \let\gplgaddtomacro\g@addto@macro
  \gdef\gplbacktext{}%
  \gdef\gplfronttext{}%
  \makeatother
  \ifGPblacktext
    \def\colorrgb#1{}%
    \def\colorgray#1{}%
  \else
    \ifGPcolor
      \def\colorrgb#1{\color[rgb]{#1}}%
      \def\colorgray#1{\color[gray]{#1}}%
      \expandafter\def\csname LTw\endcsname{\color{white}}%
      \expandafter\def\csname LTb\endcsname{\color{black}}%
      \expandafter\def\csname LTa\endcsname{\color{black}}%
      \expandafter\def\csname LT0\endcsname{\color[rgb]{1,0,0}}%
      \expandafter\def\csname LT1\endcsname{\color[rgb]{0,1,0}}%
      \expandafter\def\csname LT2\endcsname{\color[rgb]{0,0,1}}%
      \expandafter\def\csname LT3\endcsname{\color[rgb]{1,0,1}}%
      \expandafter\def\csname LT4\endcsname{\color[rgb]{0,1,1}}%
      \expandafter\def\csname LT5\endcsname{\color[rgb]{1,1,0}}%
      \expandafter\def\csname LT6\endcsname{\color[rgb]{0,0,0}}%
      \expandafter\def\csname LT7\endcsname{\color[rgb]{1,0.3,0}}%
      \expandafter\def\csname LT8\endcsname{\color[rgb]{0.5,0.5,0.5}}%
    \else
      \def\colorrgb#1{\color{black}}%
      \def\colorgray#1{\color[gray]{#1}}%
      \expandafter\def\csname LTw\endcsname{\color{white}}%
      \expandafter\def\csname LTb\endcsname{\color{black}}%
      \expandafter\def\csname LTa\endcsname{\color{black}}%
      \expandafter\def\csname LT0\endcsname{\color{black}}%
      \expandafter\def\csname LT1\endcsname{\color{black}}%
      \expandafter\def\csname LT2\endcsname{\color{black}}%
      \expandafter\def\csname LT3\endcsname{\color{black}}%
      \expandafter\def\csname LT4\endcsname{\color{black}}%
      \expandafter\def\csname LT5\endcsname{\color{black}}%
      \expandafter\def\csname LT6\endcsname{\color{black}}%
      \expandafter\def\csname LT7\endcsname{\color{black}}%
      \expandafter\def\csname LT8\endcsname{\color{black}}%
    \fi
  \fi
    \setlength{\unitlength}{0.0500bp}%
    \ifx\gptboxheight\undefined%
      \newlength{\gptboxheight}%
      \newlength{\gptboxwidth}%
      \newsavebox{\gptboxtext}%
    \fi%
    \setlength{\fboxrule}{0.5pt}%
    \setlength{\fboxsep}{1pt}%
\begin{picture}(3456.00,2590.00)%
    \gplgaddtomacro\gplbacktext{%
      \csname LTb\endcsname%
      \put(264,440){\makebox(0,0)[r]{\strut{}$-3$}}%
      \csname LTb\endcsname%
      \put(264,707){\makebox(0,0)[r]{\strut{}$-2$}}%
      \csname LTb\endcsname%
      \put(264,973){\makebox(0,0)[r]{\strut{}$-1$}}%
      \csname LTb\endcsname%
      \put(264,1240){\makebox(0,0)[r]{\strut{}$0$}}%
      \csname LTb\endcsname%
      \put(264,1506){\makebox(0,0)[r]{\strut{}$1$}}%
      \csname LTb\endcsname%
      \put(264,1773){\makebox(0,0)[r]{\strut{}$2$}}%
      \csname LTb\endcsname%
      \put(264,2039){\makebox(0,0)[r]{\strut{}$3$}}%
      \put(396,220){\makebox(0,0){\strut{}1970}}%
      \put(929,220){\makebox(0,0){\strut{}1980}}%
      \put(1461,220){\makebox(0,0){\strut{}1990}}%
      \put(1994,220){\makebox(0,0){\strut{}2000}}%
      \put(2526,220){\makebox(0,0){\strut{}2010}}%
      \put(3059,220){\makebox(0,0){\strut{}2020}}%
    }%
    \gplgaddtomacro\gplfronttext{%
      \csname LTb\endcsname%
      \put(-22,1239){\rotatebox{-270}{\makebox(0,0){\strut{}}}}%
      \put(1727,2259){\makebox(0,0){\strut{}$\Delta X$, mas (C04 series)}}%
      \csname LTb\endcsname%
      \put(2072,833){\makebox(0,0)[r]{\strut{}DE tides}}%
      \csname LTb\endcsname%
      \put(2072,613){\makebox(0,0)[r]{\strut{}IERS tides}}%
    }%
    \gplbacktext
    \put(0,0){\includegraphics{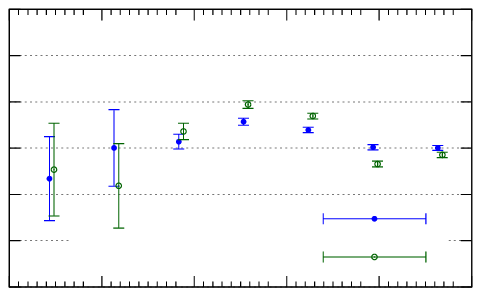}}%
    \gplfronttext
  \end{picture}%
\endgroup

%% file: icrf-x-finals.tex
\begingroup
  \makeatletter
  \providecommand\color[2][]{%
    \GenericError{(gnuplot) \space\space\space\@spaces}{%
      Package color not loaded in conjunction with
      terminal option `colourtext'%
    }{See the gnuplot documentation for explanation.%
    }{Either use 'blacktext' in gnuplot or load the package
      color.sty in LaTeX.}%
    \renewcommand\color[2][]{}%
  }%
  \providecommand\includegraphics[2][]{%
    \GenericError{(gnuplot) \space\space\space\@spaces}{%
      Package graphicx or graphics not loaded%
    }{See the gnuplot documentation for explanation.%
    }{The gnuplot epslatex terminal needs graphicx.sty or graphics.sty.}%
    \renewcommand\includegraphics[2][]{}%
  }%
  \providecommand\rotatebox[2]{#2}%
  \@ifundefined{ifGPcolor}{%
    \newif\ifGPcolor
    \GPcolorfalse
  }{}%
  \@ifundefined{ifGPblacktext}{%
    \newif\ifGPblacktext
    \GPblacktexttrue
  }{}%
  \let\gplgaddtomacro\g@addto@macro
  \gdef\gplbacktext{}%
  \gdef\gplfronttext{}%
  \makeatother
  \ifGPblacktext
    \def\colorrgb#1{}%
    \def\colorgray#1{}%
  \else
    \ifGPcolor
      \def\colorrgb#1{\color[rgb]{#1}}%
      \def\colorgray#1{\color[gray]{#1}}%
      \expandafter\def\csname LTw\endcsname{\color{white}}%
      \expandafter\def\csname LTb\endcsname{\color{black}}%
      \expandafter\def\csname LTa\endcsname{\color{black}}%
      \expandafter\def\csname LT0\endcsname{\color[rgb]{1,0,0}}%
      \expandafter\def\csname LT1\endcsname{\color[rgb]{0,1,0}}%
      \expandafter\def\csname LT2\endcsname{\color[rgb]{0,0,1}}%
      \expandafter\def\csname LT3\endcsname{\color[rgb]{1,0,1}}%
      \expandafter\def\csname LT4\endcsname{\color[rgb]{0,1,1}}%
      \expandafter\def\csname LT5\endcsname{\color[rgb]{1,1,0}}%
      \expandafter\def\csname LT6\endcsname{\color[rgb]{0,0,0}}%
      \expandafter\def\csname LT7\endcsname{\color[rgb]{1,0.3,0}}%
      \expandafter\def\csname LT8\endcsname{\color[rgb]{0.5,0.5,0.5}}%
    \else
      \def\colorrgb#1{\color{black}}%
      \def\colorgray#1{\color[gray]{#1}}%
      \expandafter\def\csname LTw\endcsname{\color{white}}%
      \expandafter\def\csname LTb\endcsname{\color{black}}%
      \expandafter\def\csname LTa\endcsname{\color{black}}%
      \expandafter\def\csname LT0\endcsname{\color{black}}%
      \expandafter\def\csname LT1\endcsname{\color{black}}%
      \expandafter\def\csname LT2\endcsname{\color{black}}%
      \expandafter\def\csname LT3\endcsname{\color{black}}%
      \expandafter\def\csname LT4\endcsname{\color{black}}%
      \expandafter\def\csname LT5\endcsname{\color{black}}%
      \expandafter\def\csname LT6\endcsname{\color{black}}%
      \expandafter\def\csname LT7\endcsname{\color{black}}%
      \expandafter\def\csname LT8\endcsname{\color{black}}%
    \fi
  \fi
    \setlength{\unitlength}{0.0500bp}%
    \ifx\gptboxheight\undefined%
      \newlength{\gptboxheight}%
      \newlength{\gptboxwidth}%
      \newsavebox{\gptboxtext}%
    \fi%
    \setlength{\fboxrule}{0.5pt}%
    \setlength{\fboxsep}{1pt}%
\begin{picture}(3456.00,2590.00)%
    \gplgaddtomacro\gplbacktext{%
      \csname LTb\endcsname%
      \put(264,440){\makebox(0,0)[r]{\strut{}$-3$}}%
      \csname LTb\endcsname%
      \put(264,707){\makebox(0,0)[r]{\strut{}$-2$}}%
      \csname LTb\endcsname%
      \put(264,973){\makebox(0,0)[r]{\strut{}$-1$}}%
      \csname LTb\endcsname%
      \put(264,1240){\makebox(0,0)[r]{\strut{}$0$}}%
      \csname LTb\endcsname%
      \put(264,1506){\makebox(0,0)[r]{\strut{}$1$}}%
      \csname LTb\endcsname%
      \put(264,1773){\makebox(0,0)[r]{\strut{}$2$}}%
      \csname LTb\endcsname%
      \put(264,2039){\makebox(0,0)[r]{\strut{}$3$}}%
      \put(396,220){\makebox(0,0){\strut{}1970}}%
      \put(929,220){\makebox(0,0){\strut{}1980}}%
      \put(1461,220){\makebox(0,0){\strut{}1990}}%
      \put(1994,220){\makebox(0,0){\strut{}2000}}%
      \put(2526,220){\makebox(0,0){\strut{}2010}}%
      \put(3059,220){\makebox(0,0){\strut{}2020}}%
    }%
    \gplgaddtomacro\gplfronttext{%
      \csname LTb\endcsname%
      \put(-22,1239){\rotatebox{-270}{\makebox(0,0){\strut{}}}}%
      \put(1727,2259){\makebox(0,0){\strut{}$\Delta X$, mas (``finals'' series)}}%
      \csname LTb\endcsname%
      \put(2072,833){\makebox(0,0)[r]{\strut{}DE tides}}%
      \csname LTb\endcsname%
      \put(2072,613){\makebox(0,0)[r]{\strut{}IERS tides}}%
    }%
    \gplbacktext
    \put(0,0){\includegraphics{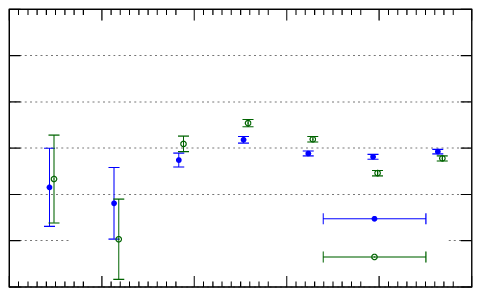}}%
    \gplfronttext
  \end{picture}%
\endgroup

%% file: icrf-y-c04.tex
\begingroup
  \makeatletter
  \providecommand\color[2][]{%
    \GenericError{(gnuplot) \space\space\space\@spaces}{%
      Package color not loaded in conjunction with
      terminal option `colourtext'%
    }{See the gnuplot documentation for explanation.%
    }{Either use 'blacktext' in gnuplot or load the package
      color.sty in LaTeX.}%
    \renewcommand\color[2][]{}%
  }%
  \providecommand\includegraphics[2][]{%
    \GenericError{(gnuplot) \space\space\space\@spaces}{%
      Package graphicx or graphics not loaded%
    }{See the gnuplot documentation for explanation.%
    }{The gnuplot epslatex terminal needs graphicx.sty or graphics.sty.}%
    \renewcommand\includegraphics[2][]{}%
  }%
  \providecommand\rotatebox[2]{#2}%
  \@ifundefined{ifGPcolor}{%
    \newif\ifGPcolor
    \GPcolorfalse
  }{}%
  \@ifundefined{ifGPblacktext}{%
    \newif\ifGPblacktext
    \GPblacktexttrue
  }{}%
  \let\gplgaddtomacro\g@addto@macro
  \gdef\gplbacktext{}%
  \gdef\gplfronttext{}%
  \makeatother
  \ifGPblacktext
    \def\colorrgb#1{}%
    \def\colorgray#1{}%
  \else
    \ifGPcolor
      \def\colorrgb#1{\color[rgb]{#1}}%
      \def\colorgray#1{\color[gray]{#1}}%
      \expandafter\def\csname LTw\endcsname{\color{white}}%
      \expandafter\def\csname LTb\endcsname{\color{black}}%
      \expandafter\def\csname LTa\endcsname{\color{black}}%
      \expandafter\def\csname LT0\endcsname{\color[rgb]{1,0,0}}%
      \expandafter\def\csname LT1\endcsname{\color[rgb]{0,1,0}}%
      \expandafter\def\csname LT2\endcsname{\color[rgb]{0,0,1}}%
      \expandafter\def\csname LT3\endcsname{\color[rgb]{1,0,1}}%
      \expandafter\def\csname LT4\endcsname{\color[rgb]{0,1,1}}%
      \expandafter\def\csname LT5\endcsname{\color[rgb]{1,1,0}}%
      \expandafter\def\csname LT6\endcsname{\color[rgb]{0,0,0}}%
      \expandafter\def\csname LT7\endcsname{\color[rgb]{1,0.3,0}}%
      \expandafter\def\csname LT8\endcsname{\color[rgb]{0.5,0.5,0.5}}%
    \else
      \def\colorrgb#1{\color{black}}%
      \def\colorgray#1{\color[gray]{#1}}%
      \expandafter\def\csname LTw\endcsname{\color{white}}%
      \expandafter\def\csname LTb\endcsname{\color{black}}%
      \expandafter\def\csname LTa\endcsname{\color{black}}%
      \expandafter\def\csname LT0\endcsname{\color{black}}%
      \expandafter\def\csname LT1\endcsname{\color{black}}%
      \expandafter\def\csname LT2\endcsname{\color{black}}%
      \expandafter\def\csname LT3\endcsname{\color{black}}%
      \expandafter\def\csname LT4\endcsname{\color{black}}%
      \expandafter\def\csname LT5\endcsname{\color{black}}%
      \expandafter\def\csname LT6\endcsname{\color{black}}%
      \expandafter\def\csname LT7\endcsname{\color{black}}%
      \expandafter\def\csname LT8\endcsname{\color{black}}%
    \fi
  \fi
    \setlength{\unitlength}{0.0500bp}%
    \ifx\gptboxheight\undefined%
      \newlength{\gptboxheight}%
      \newlength{\gptboxwidth}%
      \newsavebox{\gptboxtext}%
    \fi%
    \setlength{\fboxrule}{0.5pt}%
    \setlength{\fboxsep}{1pt}%
\begin{picture}(3456.00,2590.00)%
    \gplgaddtomacro\gplbacktext{%
      \csname LTb\endcsname%
      \put(264,440){\makebox(0,0)[r]{\strut{}$-3$}}%
      \csname LTb\endcsname%
      \put(264,707){\makebox(0,0)[r]{\strut{}$-2$}}%
      \csname LTb\endcsname%
      \put(264,973){\makebox(0,0)[r]{\strut{}$-1$}}%
      \csname LTb\endcsname%
      \put(264,1240){\makebox(0,0)[r]{\strut{}$0$}}%
      \csname LTb\endcsname%
      \put(264,1506){\makebox(0,0)[r]{\strut{}$1$}}%
      \csname LTb\endcsname%
      \put(264,1773){\makebox(0,0)[r]{\strut{}$2$}}%
      \csname LTb\endcsname%
      \put(264,2039){\makebox(0,0)[r]{\strut{}$3$}}%
      \put(396,220){\makebox(0,0){\strut{}1970}}%
      \put(929,220){\makebox(0,0){\strut{}1980}}%
      \put(1461,220){\makebox(0,0){\strut{}1990}}%
      \put(1994,220){\makebox(0,0){\strut{}2000}}%
      \put(2526,220){\makebox(0,0){\strut{}2010}}%
      \put(3059,220){\makebox(0,0){\strut{}2020}}%
    }%
    \gplgaddtomacro\gplfronttext{%
      \csname LTb\endcsname%
      \put(-22,1239){\rotatebox{-270}{\makebox(0,0){\strut{}}}}%
      \put(1727,2259){\makebox(0,0){\strut{}$\Delta Y$, mas (C04 series)}}%
      \csname LTb\endcsname%
      \put(2072,1866){\makebox(0,0)[r]{\strut{}DE tides}}%
      \csname LTb\endcsname%
      \put(2072,1646){\makebox(0,0)[r]{\strut{}IERS tides}}%
    }%
    \gplbacktext
    \put(0,0){\includegraphics{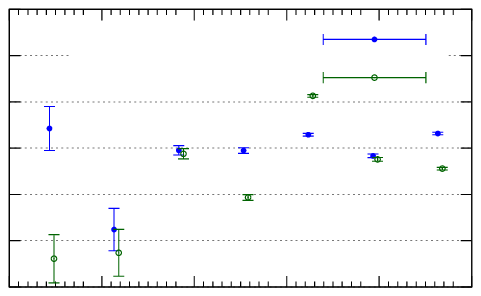}}%
    \gplfronttext
  \end{picture}%
\endgroup

%% file: icrf-y-finals.tex
\begingroup
  \makeatletter
  \providecommand\color[2][]{%
    \GenericError{(gnuplot) \space\space\space\@spaces}{%
      Package color not loaded in conjunction with
      terminal option `colourtext'%
    }{See the gnuplot documentation for explanation.%
    }{Either use 'blacktext' in gnuplot or load the package
      color.sty in LaTeX.}%
    \renewcommand\color[2][]{}%
  }%
  \providecommand\includegraphics[2][]{%
    \GenericError{(gnuplot) \space\space\space\@spaces}{%
      Package graphicx or graphics not loaded%
    }{See the gnuplot documentation for explanation.%
    }{The gnuplot epslatex terminal needs graphicx.sty or graphics.sty.}%
    \renewcommand\includegraphics[2][]{}%
  }%
  \providecommand\rotatebox[2]{#2}%
  \@ifundefined{ifGPcolor}{%
    \newif\ifGPcolor
    \GPcolorfalse
  }{}%
  \@ifundefined{ifGPblacktext}{%
    \newif\ifGPblacktext
    \GPblacktexttrue
  }{}%
  \let\gplgaddtomacro\g@addto@macro
  \gdef\gplbacktext{}%
  \gdef\gplfronttext{}%
  \makeatother
  \ifGPblacktext
    \def\colorrgb#1{}%
    \def\colorgray#1{}%
  \else
    \ifGPcolor
      \def\colorrgb#1{\color[rgb]{#1}}%
      \def\colorgray#1{\color[gray]{#1}}%
      \expandafter\def\csname LTw\endcsname{\color{white}}%
      \expandafter\def\csname LTb\endcsname{\color{black}}%
      \expandafter\def\csname LTa\endcsname{\color{black}}%
      \expandafter\def\csname LT0\endcsname{\color[rgb]{1,0,0}}%
      \expandafter\def\csname LT1\endcsname{\color[rgb]{0,1,0}}%
      \expandafter\def\csname LT2\endcsname{\color[rgb]{0,0,1}}%
      \expandafter\def\csname LT3\endcsname{\color[rgb]{1,0,1}}%
      \expandafter\def\csname LT4\endcsname{\color[rgb]{0,1,1}}%
      \expandafter\def\csname LT5\endcsname{\color[rgb]{1,1,0}}%
      \expandafter\def\csname LT6\endcsname{\color[rgb]{0,0,0}}%
      \expandafter\def\csname LT7\endcsname{\color[rgb]{1,0.3,0}}%
      \expandafter\def\csname LT8\endcsname{\color[rgb]{0.5,0.5,0.5}}%
    \else
      \def\colorrgb#1{\color{black}}%
      \def\colorgray#1{\color[gray]{#1}}%
      \expandafter\def\csname LTw\endcsname{\color{white}}%
      \expandafter\def\csname LTb\endcsname{\color{black}}%
      \expandafter\def\csname LTa\endcsname{\color{black}}%
      \expandafter\def\csname LT0\endcsname{\color{black}}%
      \expandafter\def\csname LT1\endcsname{\color{black}}%
      \expandafter\def\csname LT2\endcsname{\color{black}}%
      \expandafter\def\csname LT3\endcsname{\color{black}}%
      \expandafter\def\csname LT4\endcsname{\color{black}}%
      \expandafter\def\csname LT5\endcsname{\color{black}}%
      \expandafter\def\csname LT6\endcsname{\color{black}}%
      \expandafter\def\csname LT7\endcsname{\color{black}}%
      \expandafter\def\csname LT8\endcsname{\color{black}}%
    \fi
  \fi
    \setlength{\unitlength}{0.0500bp}%
    \ifx\gptboxheight\undefined%
      \newlength{\gptboxheight}%
      \newlength{\gptboxwidth}%
      \newsavebox{\gptboxtext}%
    \fi%
    \setlength{\fboxrule}{0.5pt}%
    \setlength{\fboxsep}{1pt}%
\begin{picture}(3456.00,2590.00)%
    \gplgaddtomacro\gplbacktext{%
      \csname LTb\endcsname%
      \put(264,440){\makebox(0,0)[r]{\strut{}$-3$}}%
      \csname LTb\endcsname%
      \put(264,707){\makebox(0,0)[r]{\strut{}$-2$}}%
      \csname LTb\endcsname%
      \put(264,973){\makebox(0,0)[r]{\strut{}$-1$}}%
      \csname LTb\endcsname%
      \put(264,1240){\makebox(0,0)[r]{\strut{}$0$}}%
      \csname LTb\endcsname%
      \put(264,1506){\makebox(0,0)[r]{\strut{}$1$}}%
      \csname LTb\endcsname%
      \put(264,1773){\makebox(0,0)[r]{\strut{}$2$}}%
      \csname LTb\endcsname%
      \put(264,2039){\makebox(0,0)[r]{\strut{}$3$}}%
      \put(396,220){\makebox(0,0){\strut{}1970}}%
      \put(929,220){\makebox(0,0){\strut{}1980}}%
      \put(1461,220){\makebox(0,0){\strut{}1990}}%
      \put(1994,220){\makebox(0,0){\strut{}2000}}%
      \put(2526,220){\makebox(0,0){\strut{}2010}}%
      \put(3059,220){\makebox(0,0){\strut{}2020}}%
    }%
    \gplgaddtomacro\gplfronttext{%
      \csname LTb\endcsname%
      \put(-22,1239){\rotatebox{-270}{\makebox(0,0){\strut{}}}}%
      \put(1727,2259){\makebox(0,0){\strut{}$\Delta Y$, mas (``finals'' series)}}%
      \csname LTb\endcsname%
      \put(2072,1866){\makebox(0,0)[r]{\strut{}DE tides}}%
      \csname LTb\endcsname%
      \put(2072,1646){\makebox(0,0)[r]{\strut{}IERS tides}}%
    }%
    \gplbacktext
    \put(0,0){\includegraphics{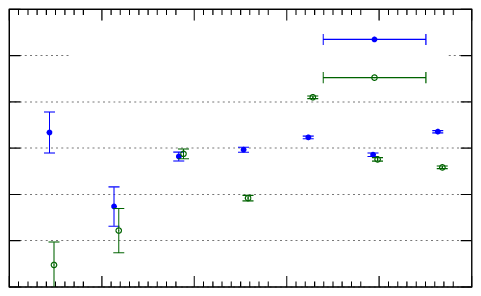}}%
    \gplfronttext
  \end{picture}%
\endgroup